\documentclass[apj]{emulateapj}
\usepackage{natbib}
\usepackage{color}
\usepackage{graphicx}
\bibpunct{(}{)}{;}{a}{}{,}
\begin{document}

   \title{
%Kinematics of Selected Local Extremely Metal Poor Galaxies  
Kinematics of Extremely Metal-poor Galaxies: \\
Evidence for Stellar Feedback
}
\author{A. Olmo-Garc\'\i a\altaffilmark{1,2},
J. S\'anchez Almeida\altaffilmark{1,2},
C. Mu\~noz-Tu\~n\'on\altaffilmark{1,2},\\
M.~E.~Filho\altaffilmark{1,2,3},
B.~G.~Elmegreen\altaffilmark{4}, 
D.~M.~Elmegreen\altaffilmark{5},
E. P\'erez-Montero\altaffilmark{6}, 
%R.~Amor\'\i n\altaffilmark{6},\\
J.~M\'endez-Abreu\altaffilmark{7}\\
%-- NO Y.~Ascasibar\altaffilmark{10},
%-- NO P.~Papaderos\altaffilmark{8,9},\\
%-- NO J.~M.~V\'\i lchez\altaffilmark{5}
}

\altaffiltext{1}{Instituto Astrof\'\i sica de Canarias, 38200 La Laguna, Tenerife, Spain}
\altaffiltext{2}{Departamento de Astrof\'\i sica, Universidad de La Laguna} 
\altaffiltext{3}{SIM/FEUP, 4200-465 Porto, Portugal}
\altaffiltext{4}{IBM Research Division, T.J. Watson Research Center, Yorktown Heights, NY 10598, USA}
\altaffiltext{5}{Department of Physics and Astronomy, Vassar College, Poughkeepsie, NY 12604, USA}
\altaffiltext{6}{Instituto de Astrof\'\i sica de Andaluc\'\i a,  CSIC, Granada, Spain}
%\altaffiltext{6}{INAF-Osservatorio Astronomico di Roma, Monte Porzio Catone, Italy}
\altaffiltext{7}{School of Physics and Astronomy, University of St Andrews, St Andrews, UK}
%\altaffiltext{10}{Universidad AutÃ³noma de Madrid, Madrid, Spain}
%
% 
\email{jos@iac.es}
\begin{abstract}
The extremely metal-poor (XMP) galaxies analyzed in a previous paper have
large star-forming regions with a metallicity lower than the rest of the galaxy. 
Such a chemical inhomogeneity reveals the external origin of the metal-poor gas fueling 
star formation, possibly indicating accretion from the cosmic web. This paper studies the 
kinematic properties of the ionized gas in these galaxies. 
Most XMPs have rotation velocity around a few tens of km\,s$^{-1}$.
The star-forming regions appear to move coherently.
%be kinematically distinct entities.
The velocity is constant within each region, and the velocity dispersion sometimes increases 
within the star-forming clump towards the galaxy midpoint, suggesting 
inspiral motion toward the galaxy center. 
Other regions present a local maximum in velocity dispersion at their center,
suggesting a moderate global expansion.
%\modified{Local minima in the velocity dispersion are observed too.}
%
% Contrary to the metallicity, the N/O ratio remains constant along the galaxy.
%, as expected in the metal-poor gas accretion scenario. 
%
The H$\alpha$ line wings show a number of faint emission features
with amplitudes around a few percent of the main H$\alpha$ component, and 
%Their amplitudes are typically a few percent of the main H$\alpha$ component, with 
wavelength shifts between 100 and 400~km\,s$^{-1}$. The components are often paired, so 
that red and blue emission features with similar amplitudes and shifts appear simultaneously. 
Assuming the faint emission to be produced by expanding shell-like structures, 
the inferred mass loading factor (mass loss rate divided by star formation rate) exceeds 10.
%a mass loss rate between $10^{-2}$ and 1~${\rm M}_\odot\,{\rm yr}^{-1}$.  The , as expected from 
%numerical simulations of galaxies that include stellar feedback.
Since the expansion velocity %of the faint emission features
exceeds by far the rotational and turbulent velocities, the gas %involved in the  expansion 
may eventually escape from the galaxy disk.
%, reaching the circum-galactic medium (CGM) and, 
%maybe, the intergalactic medium (IGM). 
%The H column density to be expected when  the shell material arrives at the CGM
%is undetectable small with the present technology.  
The observed motions involve energies consistent with
the kinetic energy 
released by individual core-collapse supernovae. 
Alternative explanations for the faint emission have been considered and discarded. 
%
%The slow global expansion of the star-forming region involves as much mass 
%loss and energy as the fast-moving small-scale expanding shells. 
%%%%
\end{abstract}
\slugcomment{this is the overleaf version}

%\shorttitle{]
% \shortauthors{}

%\tableofcontents

\section{Introduction}\label{introduction}
Galaxies in which the star-forming gas has  a metallicity smaller than a tenth of the solar metallicity are 
known as Extremely Metal-Poor galaxies \citep[XMP; e.g.,][]{2000A&ARv..10....1K}. 
Since metals are primarily produced by stars, 
the star-forming gas in XMPs is chemically primitive.
They represent less than 0.1\,\% of the objects in galaxy catalogs \citep[e.g.,][]{2016ApJ...819..110S}. 
XMPs can be divided into two types, according to their absolute magnitude.
The first type is ultra-faint late-type galaxies, with $M_B\geq -12.5$. Since galaxies
follow a well known luminosity-metallicity relation \citep[e.g.,][]{1989ApJ...347..875S},
those with $M_B\geq -12.5$ should be XMPs \citep[][]{2012ApJ...754...98B}. Such XMPs
are particularly rare in surveys \citep[e.g.,][]{2015MNRAS.448.2687J,sanchez+16b}.
The second type, with $M_B < -12.5$,
corresponds to low metallicity outliers of the luminosity-metallicity 
relation.  With high surface brightness, they dominate the catalogs of XMPs  
\citep[][]{2011ApJ...743...77M,2012A&A...546A.122I,2016ApJ...819..110S}, and 
they are the subject of our study. The evolutionary pathways of the two types probably differ. 
In this paper the term XMP refers exclusively to the second type.

%As we state above, 
In addition to having chemically primitive star-forming regions, XMPs
%They 
also show signs of being structurally primitive, as if they %XMPs 
were  disks in early stages of assembly, with massive off-center starbursts providing  
them with a characteristic tadpole shape 
\citep[e.g.,][]{2008A&A...491..113P,2011ApJ...743...77M,2012ApJ...750...95E,2016arXiv160502822E}, 
with slowly rotating bodies, and with large spatially-unresolved turbulent motions  
\citep[e.g.,][]{2013ApJ...767...74S,2015A&A...578A.105A}. 
Reinforcing such a primitive character, their chemical composition is often non-uniform
\citep[][]{2006A&A...454..119P,2009A&A...503...61I,2011ApJ...739...23L,2013ApJ...767...74S,2014ApJ...783...45S}, 
with the lowest metallicities mostly in the regions of intense star formation. The existence of chemical 
inhomogeneities is particularly revealing because the timescale 
for mixing in disk galaxies is short, of the order of a fraction
of the rotational period  \citep[e.g.,][]{2002ApJ...581.1047D,2012ApJ...758...48Y}. 
It implies that the metal-poor gas in XMPs was recently accreted from 
a nearly pristine cloud, very much in line with the expected cosmic cold-flow 
accretion %scenario 
predicted to build disk galaxies 
\citep[][]{2003MNRAS.345..349B,2005MNRAS.363....2K,2009Natur.457..451D}, 
but which has been so difficult to test observationally
\citep[e.g.,][]{2008A&ARv..15..189S,2012ARA&A..50..491P,2014A&ARv..22...71S}. 
The gas expected in cosmological gas accretion events is described in detail by 
\citet{2014A&ARv..22...71S}.
Thus, XMPs seem to be local galaxies that are growing through  
the physical process that created and shaped disk galaxies in 
the early universe. They open up a gateway to study this fundamental 
process with unprecedented detail at low redshift.

The presence of chemical inhomogeneities in most XMPs is the touchstone of 
the whole scenario, since it unequivocally shows the star formation to be feeding
from external metal-poor gas. In order to put the presence of these inhomogeneities
on a firm observational basis, we measured the oxygen abundance along 
the major axes of ten XMPs \citep[][Paper I]{2015ApJ...810L..15S}, 
using a method based on models consistent with
the direct method \citep[][]{2014MNRAS.441.2663P}. 
In nine out of the ten cases, sharp  metallicity drops were found, leaving 
little doubt  as to the existence and ubiquity of the drops.
The present paper follows up on Paper~I, and measures the kinematic properties 
of these XMPs. The spectra used  to infer metallicities in Paper~I do not 
possess enough spectral resolution to carry out kinematic studies. Therefore, 
a new set of observations was obtained. For this reason, the overlap with 
the original XMP sample is not complete. However,  most of the original galaxies are 
examined here and show evidence for the star-forming regions having coherent motions,
%kinematically decoupled components, 
which also hints at the external origin of the 
galaxy gas. In addition, the H$\alpha$ line presents multiple side lobes that seem to
trace feedback processes of the star formation on the ISM (interstellar medium)
of the galaxies.  
Since we use H$\alpha$, our measurements characterize the properties 
of the ionized gas near the sites of star formation.

The paper is organized as follows:
Sect.~\ref{observations} describes the observations and reduction, emphasizing the 
aspects required to obtain the spectral resolution needed for the kinematic analysis. 
Sect.~\ref{explaining_rules} describes how physical parameters are obtained
from the calibrated spectra. It is divided into various parts; one for the
main H$\alpha$ signal (Sect.~\ref{main}), and three for the secondary lobes
(Sects.~\ref{lobes}, \ref{mass_loss_rate}, and \ref{sec:mass_bh}).
The overall properties, such as rotation, turbulent motions, and global expansion, 
are presented in Sect.~\ref{sect:rot_turb}. In this section, we also study the uniformity of the N/O ratio.
The properties of the multiple secondary components bracketing the main H$\alpha$ component
are presented and interpreted in Sect.~\ref{multiple_comp}.
They appear to trace gas swept by
the star formation process, and in Sect.~\ref{sec:mydiscussion} 
we predict the properties when this gas reaches the CGM (circum-galactic medium).
Our results are summarized and discussed in Sect.~\ref{conclusions}.
The emission expected from an expanding dusty shell of gas is formulated in Appendix~\ref{appendix}. 

%%%%%

%%%%%%%%%%%%%%

\section{Observations and data reduction}\label{observations}

\subsection{Observations}\label{observ1}

\begin{figure*}
\includegraphics[width=0.33\textwidth,angle=0]{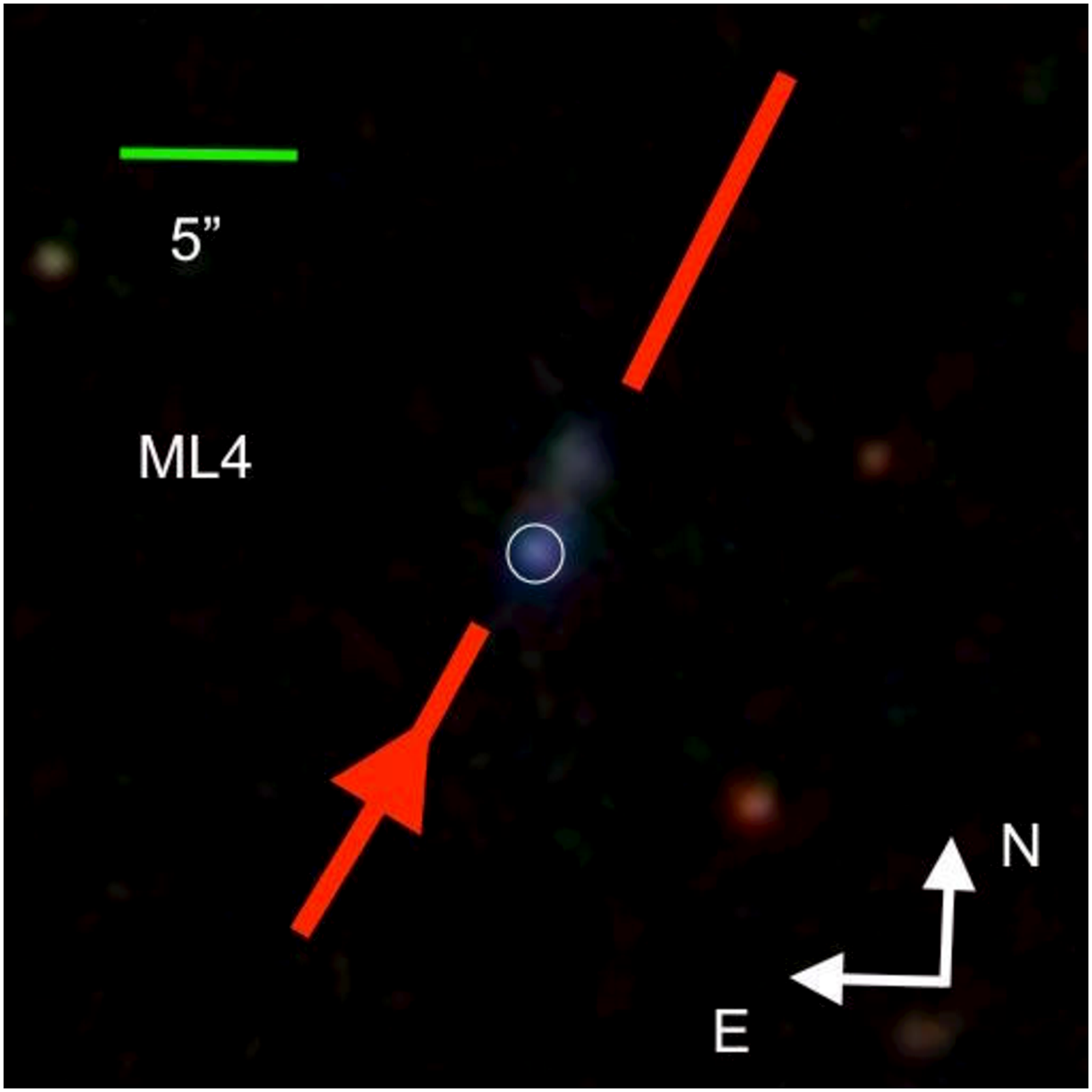} %{J0303-01.eps} 
\includegraphics[width=0.33\textwidth,angle=0]{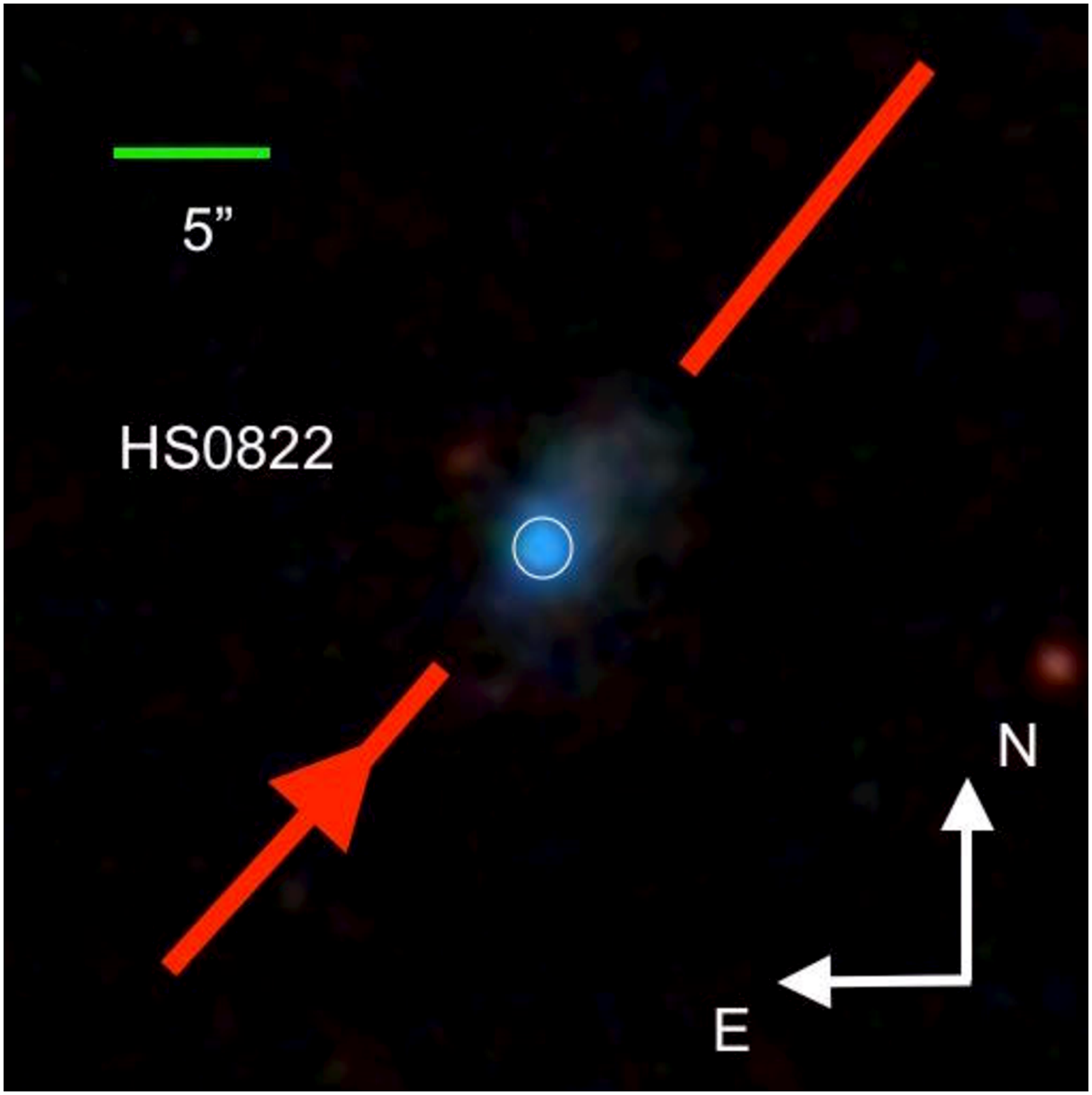} %{J0825+35.eps}
\includegraphics[width=0.33\textwidth,angle=0]{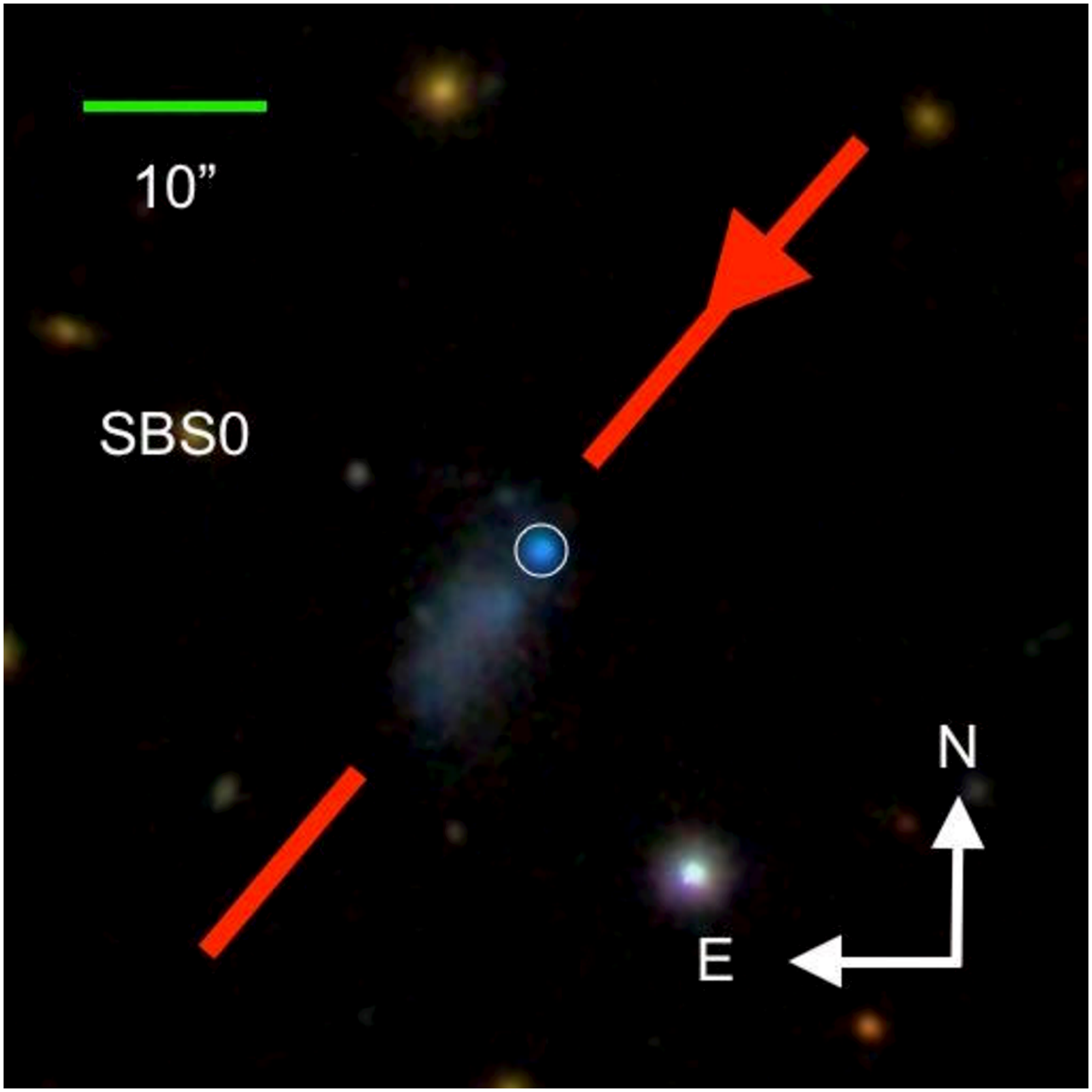} %J0944+54.eps} %{example.eps}
\includegraphics[width=0.33\textwidth,angle=0]{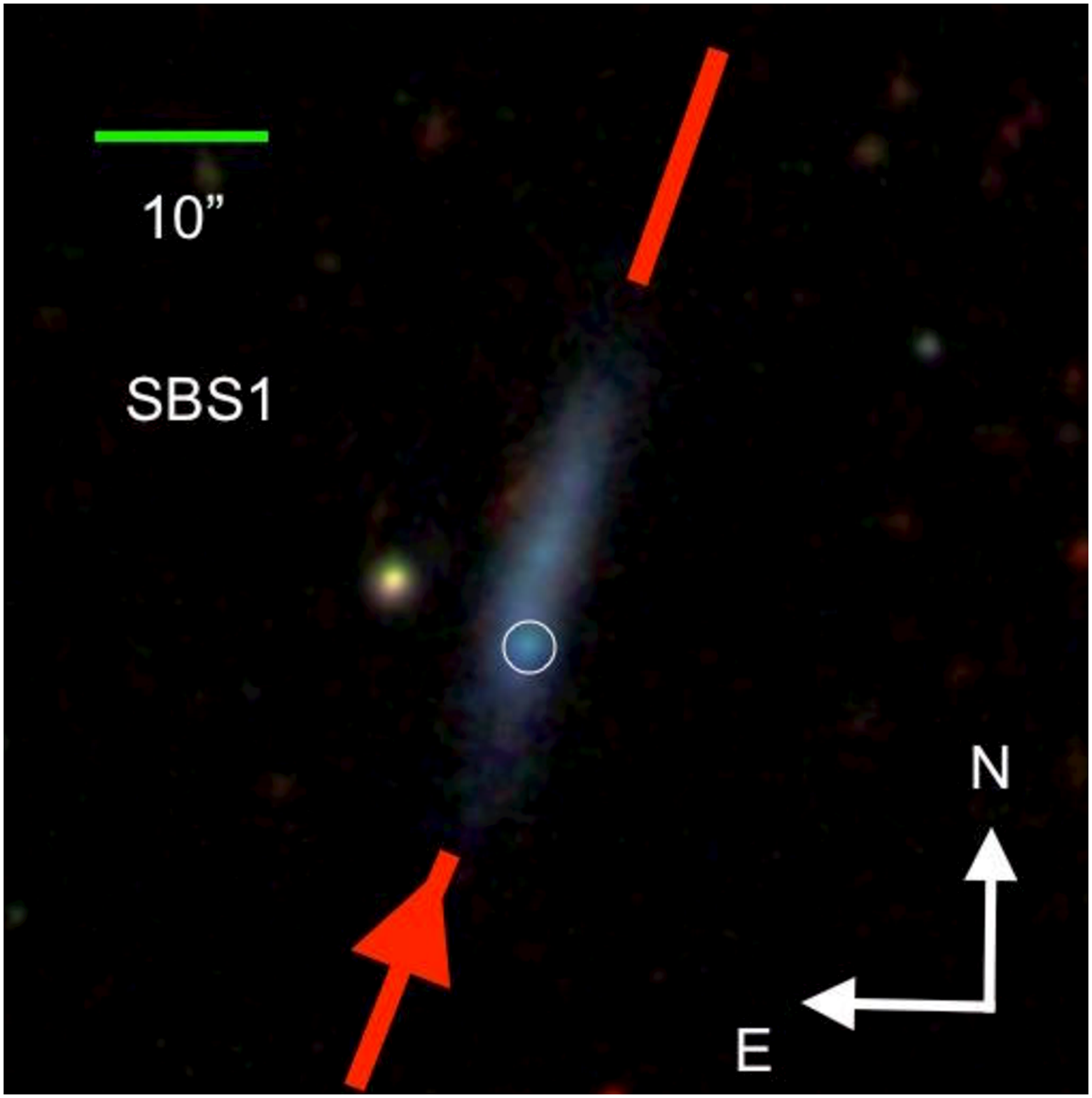} %J1132+57.eps} %{example.eps}
\includegraphics[width=0.33\textwidth,angle=0]{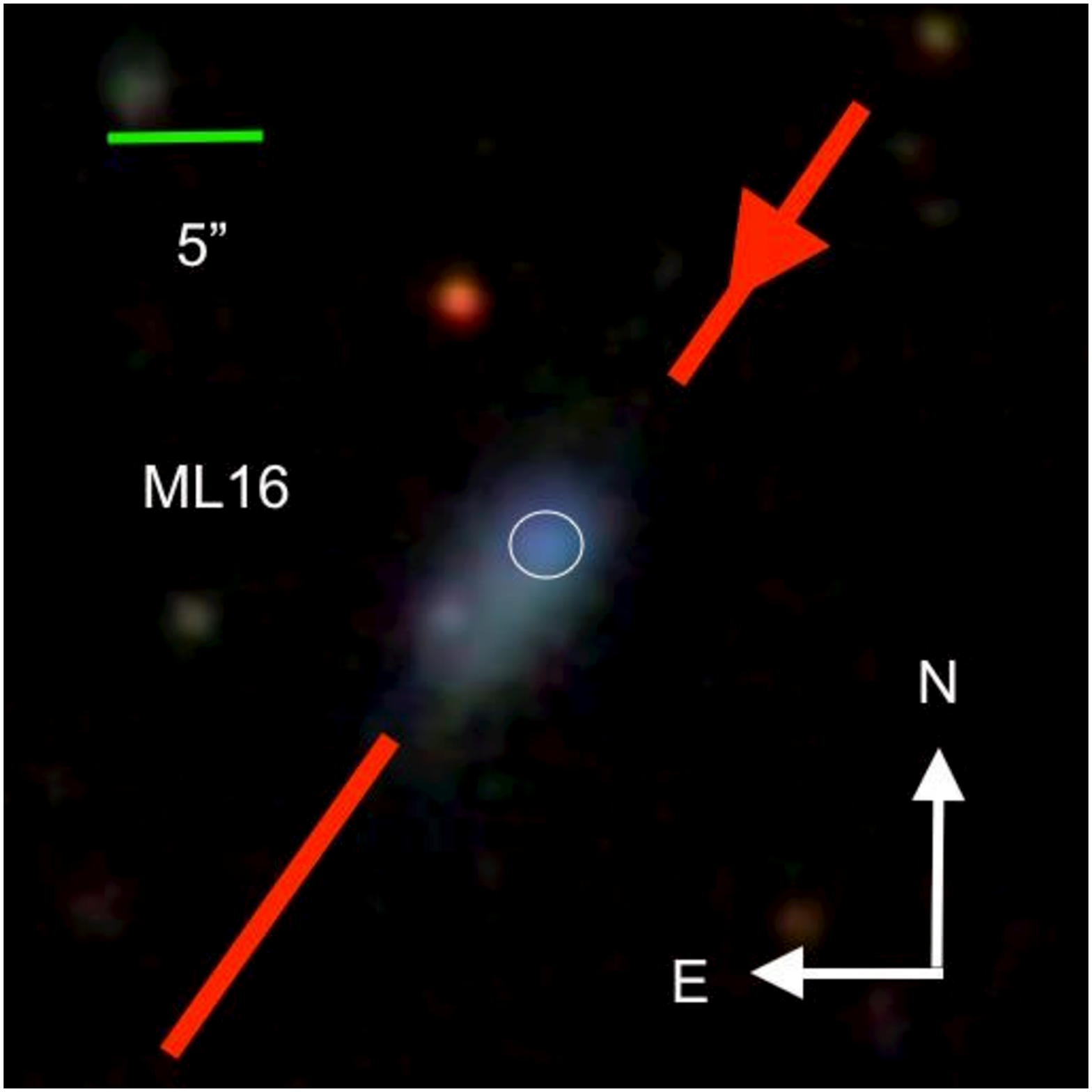} %J1145+50.eps} %{example.eps}
\includegraphics[width=0.33\textwidth,angle=0]{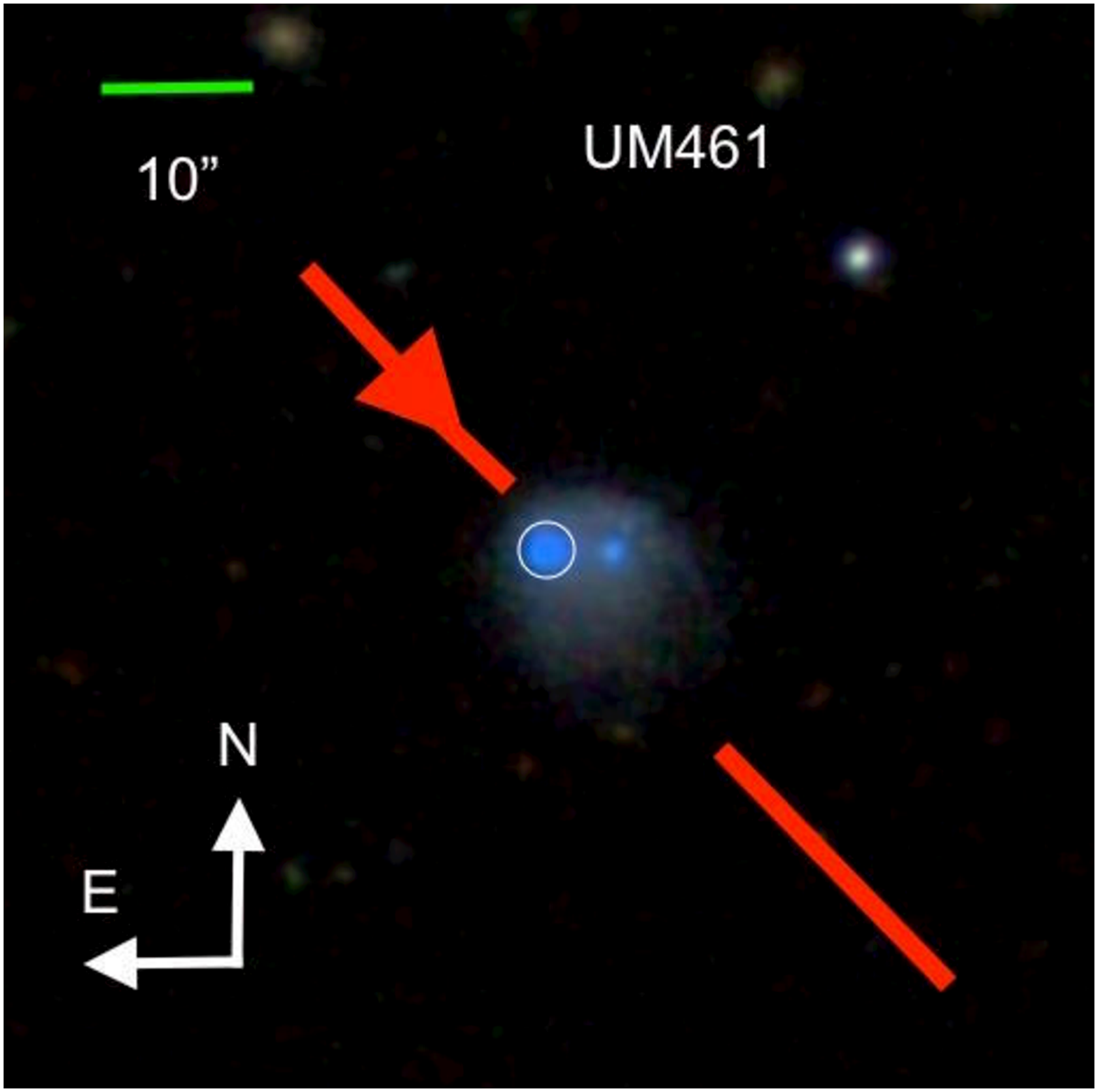} %J1151-02.eps} %{example.eps}
\includegraphics[width=0.33\textwidth,angle=0]{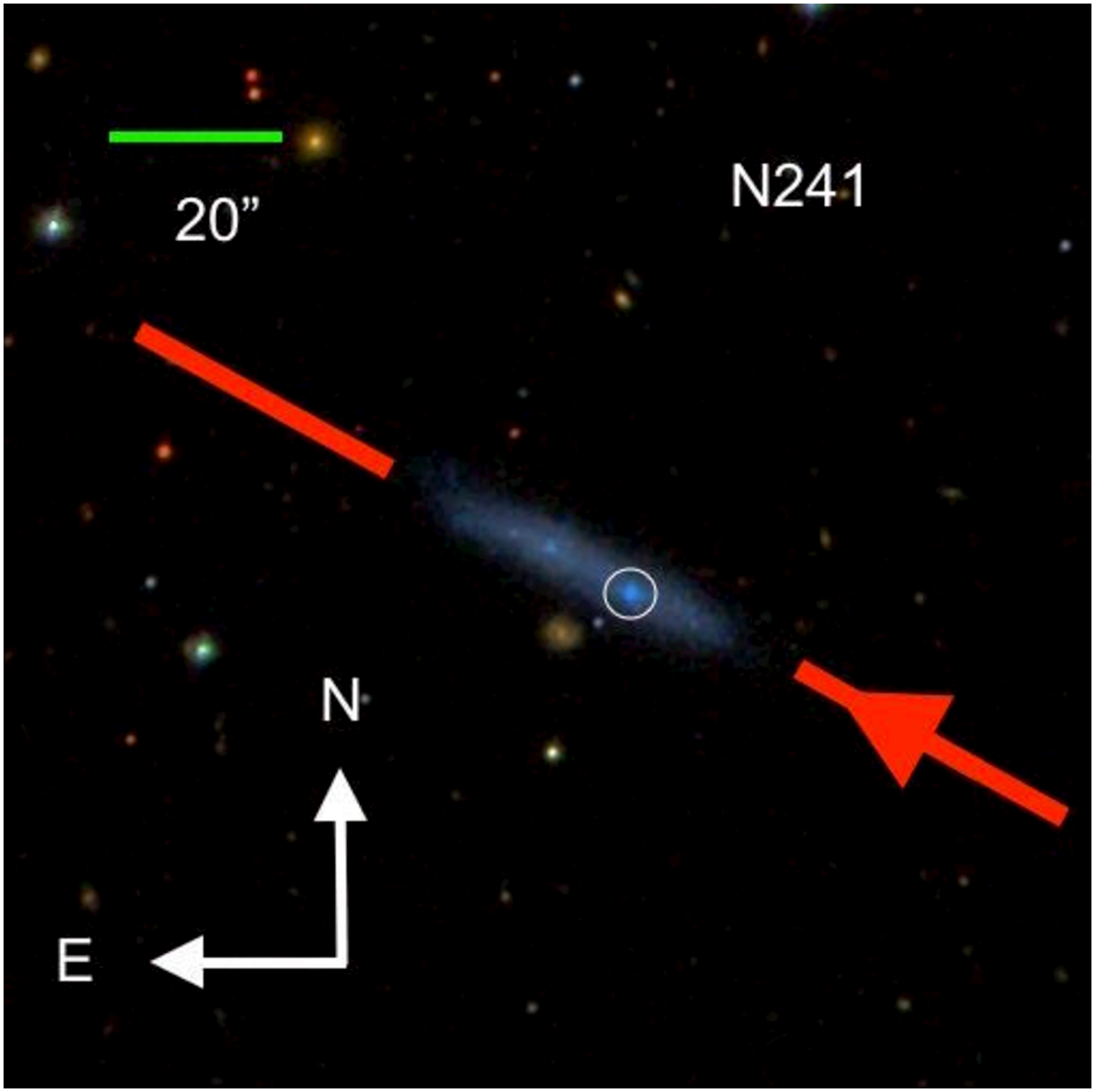} %J1444+42.eps} %{example.eps}
\includegraphics[width=0.33\textwidth,angle=0]{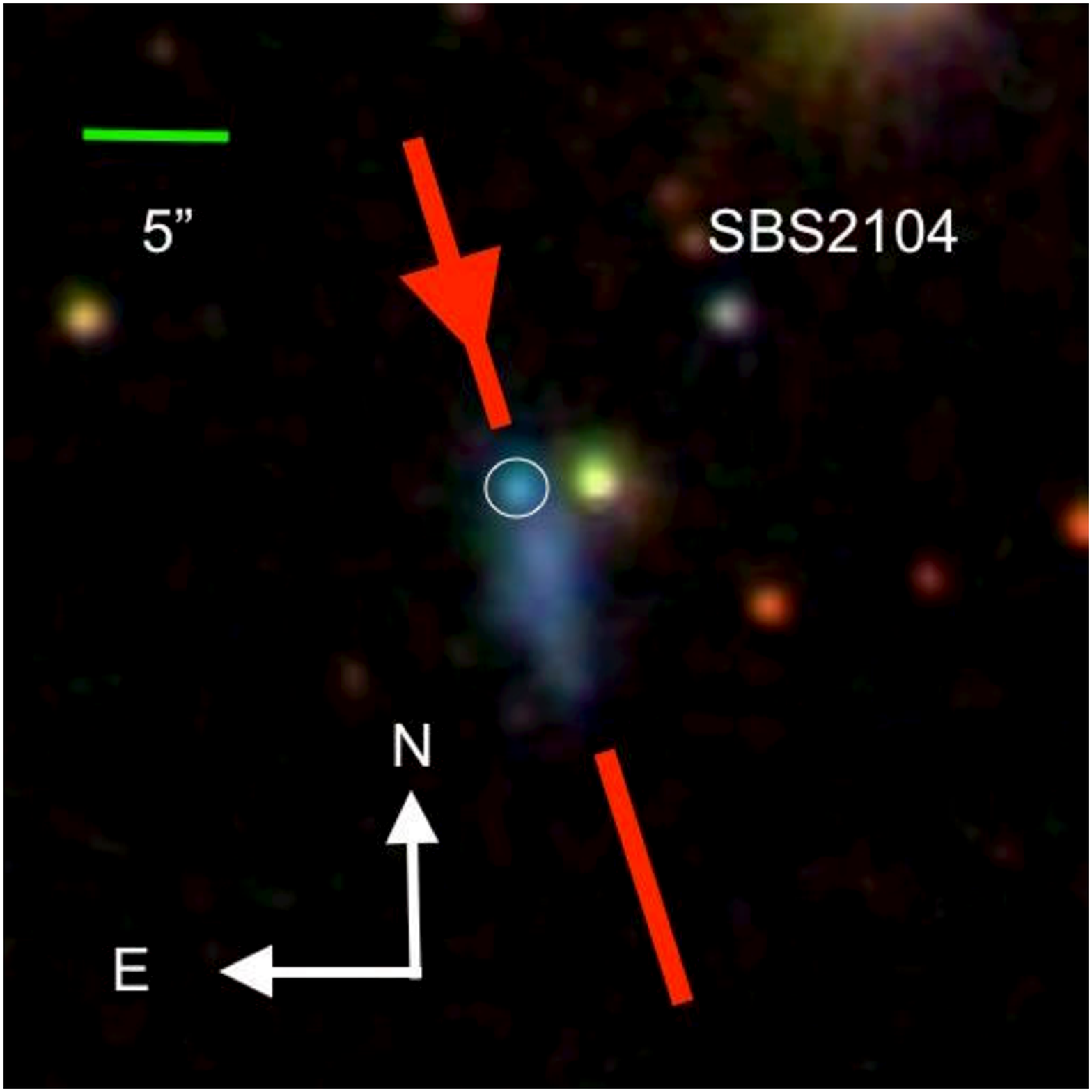} %J2104-00.eps} %{example.eps}
\includegraphics[width=0.33\textwidth,angle=0]{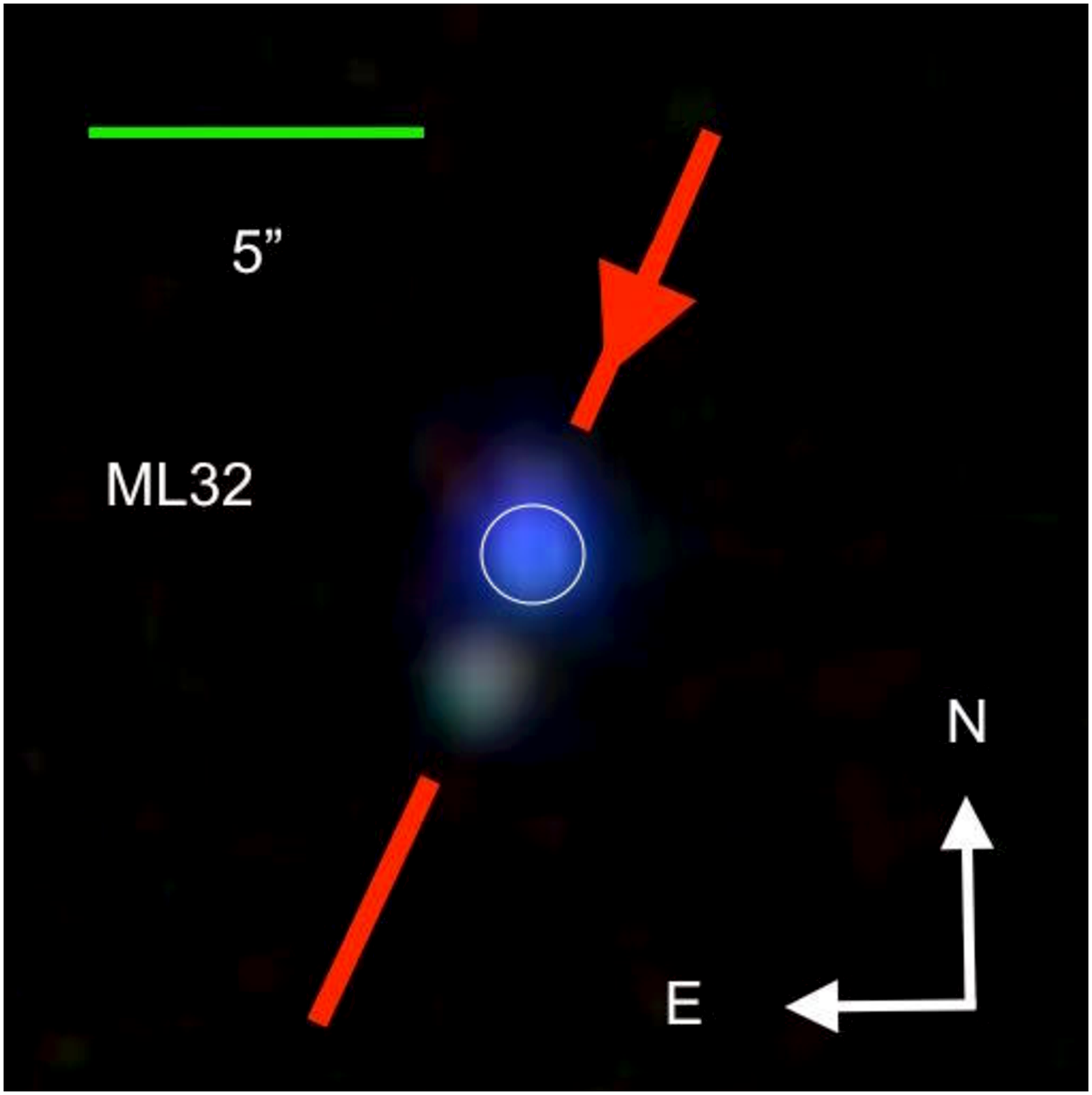} %J2302+00.eps} %{example.eps}
\caption{
SDSS color images of the galaxies observed in the paper, with the red bar 
indicating the position of the spectrograph slit, 
and the arrow pointing towards the sense of growing distance along the major axis. 
The green bar provides the scale on the sky. 
The circles enclose the main star-forming region in each galaxy.
}
\label{fig:all_images}
\end{figure*}

For the study of the kinematics of XMPs, we selected a sample of 9 galaxies, 
8 of which were previously 
studied in Paper~I (see Table~\ref{table_obs}), and thus have spatially resolved measurements of oxygen abundance. 
This time the galaxies  were observed with the ISIS\footnote{
Intermediate dispersion Spectrograph and Imaging System.
} instrument 
%\red{shoudnt we say what ISIS stands for?}
%(Intermediate dispersion Spectrograph and Imaging System) 
on the 4.2\,m William Herschel Telescope (WHT), using a high resolution
grating\footnote{{\tt http://www.ing.iac.es/astronomy/telescopes/wht/}}. 
The slit of the spectrograph was positioned along the major axis of the galaxies
(see Fig.~\ref{fig:all_images}), overlaying the slit used in Paper~I. 
%Images with the galaxies and the long-slits are given in Paper~I and will not be repeated here.
The slit width was selected to be 1 arcsec, according to the seeing at the telescope and also matching the value 
of our previous observation of these galaxies in Paper~I. The spectrograph has two arms, the
red and the blue arm, each one with its own detector. All galaxies were observed in both arms, except for HS0822, where 
we only have the red spectrum. A $2\times 2$ binning of the original images was performed to reduce oversampling.
After binning, the dispersion turns out to be 0.52~\AA\,pix$^{-1}$ for the red arm and 0.46 \AA\,pix$^{-1}$ for the blue arm.
Similarly, the spatial scale is 0.44~arcsec\,pix$^{-1}$ for the red arm and 0.40~arcsec\,pix$^{-1}$ for the blue arm.  
The wavelength coverage for the blue arm is  940~\AA\  centered at H$\beta$
(from 4390\,\AA\ to 5330\,\AA ), whereas the read arm covers 1055~\AA\ around H$\alpha$
(from 6035\,\AA\ to 7090\,\AA ). The total exposure time per target was 
approximately two hours (Table~\ref{table_obs}), divided in snapshots of 20 minutes 
as a compromise to simultaneously minimize cosmic-ray contamination and readout noise.

\begin{table*}[ht]
\caption{Parameters of the observed XMP galaxies}
\footnotesize
\centering
\begin{tabular}{lcccccccc}
\hline
\noalign{\smallskip} 
Our ID &Galaxy name & Obs. Date & Seeing\,$^b$ & $W_{Ui}$\,$^c$ & $D$\,$^d$  & Scale\,$^d$ & Exp. Time&$\log{M_\star\,^e}$\\
  &  & dd/mm/yyyy &  [arcsec] & [km\,s$^{-1}$] & [Mpc]  &  [pc\,arcsec$^{-1}$] & [s]&[M$_\odot$]\\
\noalign{\smallskip} 
\hline  
\noalign{\smallskip} 
ML4&   J$030331.26-010947.0$  &16/08/2012 & 0.5 & 38.1 $\pm$ 2.7 &  125 $\pm$ 9 & 608 & 3600 & 8.33 $\pm$ 0.33\\
HS0822&   J$082555.52+353231.9$ &02/02/2013 &1.4 & 37.8 $\pm$ 1.6 & 9.6 $\pm$ 0.7 & 47 & 7200 & 6.04 $\pm$ 0.03 \\
SBS0 &   J$094416.59+541134.3$ & 15/03/2015& 0.9& 41.1 $\pm$ 3.2 & 23.1 $\pm$ 1.6 & 112 & 4800  & 7.05 $\pm$ 0.05 \\
SBS1&   J$113202.41+572245.2$ & 15/03/2015&1.2 & 40.4 $\pm$3.6  & 22.5 $\pm$ 1.6 & 109 & 7200 & 7.53 $\pm$ 0.48\\
ML16 &   J$114506.26+501802.4$ & 15/03/2015& 1.0& 40.1 $\pm$ 4.0 & 23.6 $\pm$ 1.7 & 114 & 7200 & 6.71 $\pm$ 0.07\\
UM461 &    J$115133.34-022221.9$ & 15/03/2015& 1.2& 40.6 $\pm$ 4.0 & 12.6 $\pm$ 0.9 & 61 & 6400 & 6.7 $\pm$ 1 \\
N241$^a$&   J$144412.89+423743.6$  &15/03/2015& 1.0& 40.3 $\pm$ 4.0 &10.2 $\pm$ 0.7  & 49 & 6000 & 6.52 $\pm$ 0.07\\
SBS2104&   J$210455.31-003522.3$ &16/08/2012 & 0.5& 39.0 $\pm$ 2.2 & 20.1 $\pm$ 1.7 & 97 & 7200 & 6.19 $\pm$ 0.05 \\
ML32 &  J$230209.98+004939.0$ &16/08/2012 & 0.5 & 42 $\pm$ 5 & 138 $\pm$ 10 & 666  & 7200 & 8.39 $\pm$ 0.33\\
\noalign{\smallskip} 
\hline 
\end{tabular}
\begin{tabular}{l}
\noalign{\smallskip} 
$^a$~Target not included in Paper~I.\\
$^b$~Mean value of the seeing parameter during the observation of each object, measured by RoboDIMM@WHT~~~~~~~~~~~~~~~~\\
~~~(http://catserver.ing.iac.es/robodimm/).\\
$^c$~Mean FWHM of the telluric lines measured in each spectrum.\\
$^{d}$~Distance and scale at the position of the source, from NED (NASA/IPAC Extragalactic Database). \\ 
$^e$~Total stellar mass of the galaxy from MPA-JHU (see Paper~I).\\
\end{tabular}
\label{table_obs}
\end{table*}

The spectral resolution was measured  from the FWHM (full width at half maximum) of 
several telluric lines, using the calibrated data described in the next section. They yield  a value of approximately 40~km\,s$^{-1}$ (Table~\ref{table_obs}), which renders a spectral 
resolution close to 7500 at H$\alpha$. This resolution is in good agreement with the value expected 
for the 1~arcsec slit, once the anamorphic magnification of ISIS is taken into account 
\citep[][]{1979PASP...91..149S}.

The angular resolution during observations ranges from fair (1.3~arcsec) to 
good (0.5~arcsec), as measured by the seeing monitor at the telescope 
(Table~\ref{table_obs}). Additional information on the galaxies and 
the observations is included in Table~\ref{table_obs}. 

%%%%%%%%%%%

\subsection{Data reduction}\label{data_reduction}

\begin{figure}
\centering 
\includegraphics[width=\linewidth]{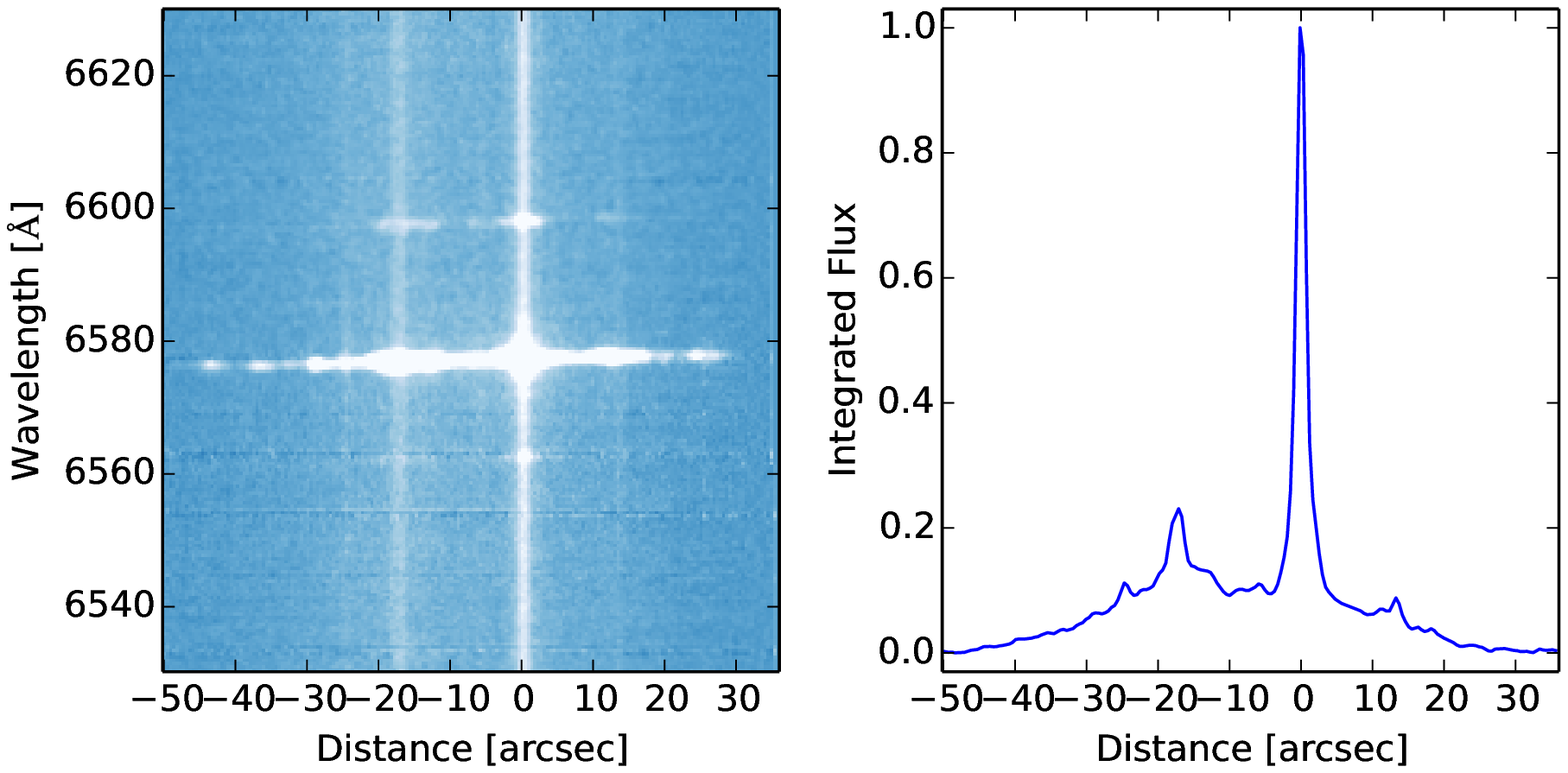}
\includegraphics[width=.95\linewidth]{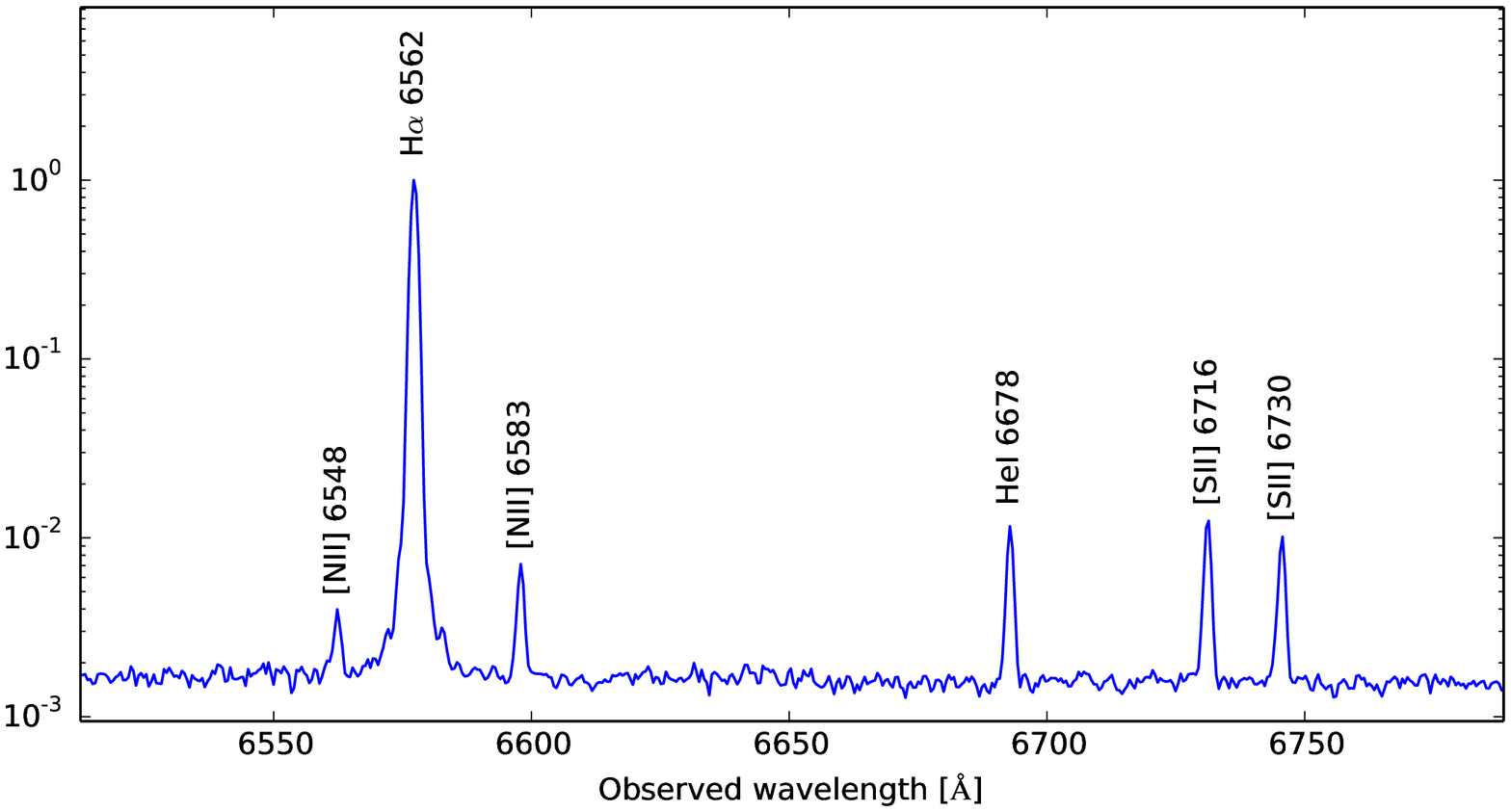}
\caption{Reduced spectrum of the galaxy N241, shown as an example.  
Top left: 2D representation of the spectrum with the observed wavelength in the vertical 
axis and the spatial position in the horizontal axis. The three emission lines that appear in the 
image are H$\alpha$ and the [NII] doublet. Top right: wavelength-integrated flux normalized to the 
maximum value. The peak emission traces the the brightest clump in the galaxy, and it defines 
the origin of the spatial scale.
Bottom: spectrum at the brightest clump, in logarithmic units 
and normalized to the peak H$\alpha$ emission. The main spectral lines used in the paper
are labeled.
}
\label{example_spect}
\end{figure}
The spectra were reduced using standard routines included in the IRAF\footnote{{\tt http://iraf.noao.edu/}~~IRAF
is  distributed  by  the  National  Optical Astronomy  Observatory,  which is  operated  by  the Association  of
Universities  for Research  in  Astronomy  (AURA) under  cooperative agreement with the National Science  Foundation.} package.  
The process comprises bias subtraction, flat-field correction using both dome and sky 
flat-fields, removal of cosmic-rays \citep{2001PASP..113.1420V}, and  
wavelength calibration using spectral lamps.
In order to evaluate the precision of the wavelength calibration, we measured 
the variation of the centroids of several sky emission lines along the spatial direction (if the calibration were error-free, the telluric line wavelength should not vary).  
For different galaxies and emission lines, the root mean square (RMS) variation of the centroids ranges 
from 0.036~\AA\ to 0.072~\AA , which corresponds to 1.6 km\,s$^{-1}$ and 3.3 km\,s$^{-1}$, respectively.
It drops even further if only strong telluric lines are considered.

The sky subtraction, the geometrical distortion correction, and the combination of the spectra of each single source were carried out 
using Python custom-made routines. 
For the sky subtraction, we performed a linear fit, sampling the sky on both sides of the galaxy spectrum. This model sky emission 
was then subtracted from each wavelength pixel. The effect of the geometrical distortion is that the spatial and wavelength 
direction are not exactly perpendicular on the detector. We determine the angle between the two directions 
as the angle between the continuum emission of the brightest point in the galaxy and the telluric lines.
A linear fit allowed us to model the geometric transformation needed to make the spatial and wavelength axes perpendicular. 
The spectra from different exposures  of the same galaxy were re-aligned before combining them.
The left panel in Fig.~\ref{example_spect} illustrates the final result of the reduction procedure.
It contains the spectrum around H$\alpha$ of the galaxy N241. 
%The resulting spectrum still contains traces from sky emission near [NII]$\lambda$6548, but they are 
%at the noise level and so are unimportant for our work. 
The signal-to-noise ratio (S/N) in continuum pixels 
has an average value of approximately 5 in the inner 10~arcsec around the brightest H{\sc ii} region of the galaxy. 
The S/N in H$\alpha$ ranges from 200 to 900 %times the continuum S/N 
at the brightest point in the galaxy,
decreasing toward the outskirts.
%
%We did not perform a detailed absolute flux calibration of the spectra because 
Because our main goal is the kinematic analysis,   
%However, 
we only performed a rough flux calibration by comparison with the calibrated Sloan Digital Sky Survey spectra of the galaxies 
\citep[SDSS DR-12; ][]{2015ApJS..219...12A}, re-scaling the flux by accounting for the difference between our 1~arcsec slit
and the 3~arcsec diameter of the SDSS fiber. Specifically, the flux integrated over the central 3~arcsec of the galaxies 
is scaled to be equal to 0.42 times the SDSS flux, where 0.42 is the ratio between the areas 
covered by the slit and the fiber. 
SDSS spectra of SBS1 and UM461 are not available. Therefore, we use for calibration the  spectra of these 
galaxies analyzed in Paper~I.

Since the photometric calibration is only approximate, we have preferred not to correct the 
spectra for internal reddening in the HII region. This correction is based on comparing the observed 
Balmer decrement  with its theoretical value \citep[e.g.,][]{1996ias..book.....E}, which
in practice implies comparing fluxes of spectra taken with the red arm (H$\alpha$) and the blue
arm (H$\beta$) of the spectrograph. Rather than carrying out an uncertain correction,
we have preferred to neglect the small expected reddening \citep[e.g.,][]{2016ApJ...819..110S}. 
If one uses the SDSS spectra of our targets to estimate the Balmer decrement, the ensuing reddening correction 
implies increasing H$\alpha$ by 50\,\% or less. Neglecting this correction does not modify any of 
the conclusions drawn in the paper.

%%%%%%%%%%%%%%%

\section{Equations used to determine physical parameters}\label{explaining_rules}

\subsection{Parameters from the main H$\alpha$ emission}\label{main}
Mean velocities, velocity dispersions, and rotation curves of the emitting gas
%, and dynamical masses 
are estimated following \citet[][Sect.~3]{2013ApJ...767...74S}, and we
refer to this work for details. For the sake of completeness, however, 
the main assumptions and the resulting equations are provided in the following. 

The velocity, $U$, was calculated from the global wavelength displacement of the H$\alpha$ emission. 
It was derived from the centroid of the line, and also from the central wavelength
of a Gaussian fitted to the emission line. Both values generally agree
with  differences $< 1$\,km\,s$^{-1}$ in the regions of interest.
The error in velocity was estimated assuming the RMS
of the continuum variations to be given by noise, and then propagating this noise into the 
estimated parameter
\citep[e.g.,][]{1971stph.book.....M}.
In the case of the centroid, the expression is analytic. For the Gaussian fit, it is 
the standard error given by the covariance matrix of the fit 
\citep[e.g.,][chapter 14]{1986nras.book.....P}.
The spatially unresolved velocity dispersion, $W_U$, 
was calculated from the width of the Gaussian model fitted to the line, 
but also directly from the line profile. We correct the observed FWHM, 
$W_{U0}$, for the instrumental spectral resolution, 
$W_{Ui}$ ($\simeq$~40~km~s$^{-1}$; see Sect.~\ref{observ1} and Table~\ref{table_obs}),
for thermal motions, $W_{Ut}$ ($\simeq$~25~km~s$^{-1}$, corresponding 
to H atoms at  $1.5\times 10^4$\,K, a temperature typical of the HII regions in our 
galaxies; see Table~\ref{tab:mean_data}), 
and for the natural width of H$\alpha$, $W_{Un}$ ($\simeq 7$~km~s$^{-1}$), so that 
\begin{equation}
W_{U}^2=W_{U0}^2-W_{Ui}^2-W_{Ut}^2-W_{Un}^2.
\label{eq:thermal}
\end{equation}
\begin{table*}[!h] 
%\scriptsize 
\caption{Physical parameters for the main star-forming region in the galaxies}
\centering 
\begin{tabular}{lccccccccc}
\hline
\noalign{\smallskip} 
ID  & HII size &  $\pounds_{H\alpha}$ & SFR & $\langle 12+\log({\rm O/H})\rangle\,^a$ & $\langle T_e\rangle\,^a$ & $\langle n_e \rangle\,^b$ & 
R$_e\,^c$ & W$_U\,^d$  & $\log{M_{\mathrm{dyn,turb}}}\,^e$  \\ 
&[arcsec]& [$10^{38}$\ erg\,s$^{-1}$] & [M$_\odot$\,yr$^{-1}$] & &[$10^4$ K]& [cm$^{-3}$] 
& [kpc] & [km s$^{-1}$] & [M$_\odot$]\\
\noalign{\smallskip} 
 \hline
 \noalign{\smallskip} 
ML4 & 1.4 &  30 $\pm$ 4 & 0.016 $\pm$ 0.002 & 7.76 $\pm$ 0.06 & 1.57 & 13 & 0.435 $\pm$ 0.012 & 62 $\pm$ 2 & 8.3 $\pm$ 0.03 \\ 
HS0822 & 1.1 &  1.9 $\pm$ 0.3 & 0.00101 $\pm$ 0.00015 & 7.69 $\pm$ 0.07 & 1.56  & 38 & 0.027 $\pm$ 0.006 & 36 $\pm$ 2  & 6.61 $\pm$ 0.1\\ 
SBS0 & 1.0 &  7.1 $\pm$ 1 & 0.0038 $\pm$ 0.0005 & 7.63 $\pm$ 0.06 & 1.67 & 129 & 0.052 $\pm$ 0.011 & 38 $\pm$ 3  & 6.95 $\pm$ 0.12\\ 
SBS1 & 1.1 & 2.3 $\pm$ 0.3 & 0.00121$\pm$ 0.00017& 7.61 $\pm$ 0.09 & 1.44 & 84 & 0.059 $\pm$ 0.018 & 51 $\pm$ 3  & 7.27 $\pm$ 0.14\\ 
ML16 & 2.4 &  7.5 $\pm$ 1.1 & 0.004 $\pm$ 0.0006 & 7.92 $\pm$ 0.04 & 1.46 & 48 & 0.139 $\pm$ 0.005 & 35 $\pm$ 5 & 7.3 $\pm$ 0.12 \\ 
UM461 & 1.2 & 1.32 $\pm$ 0.19 & 0.00071 $\pm$ 0.0001 & 7.38 $\pm$ 0.09 & 2.47 & 212 & 0.036 $\pm$ 0.009 & 42 $\pm$ 4  & 6.88 $\pm$ 0.14\\ 
N241 & 1.4 &  87 $\pm$ 13 & 0.047 $\pm$ 0.007  & 7.11 $\pm$ 0.28\,$^f$ & 1.5\,$^g$ & 73  & 0.035 $\pm$ 0.003 & 48 $\pm$ 3 & 6.99 $\pm$ 0.07\\ 
SBS2104 & 1.4 &  4.1 $\pm$ 0.7 & 0.0022 $\pm$ 0.0004 & 7.48 $\pm$ 0.08 & 1.58 & 113 & 0.069 $\pm$ 0.002 & 18 $\pm$ 5 & 6.4 $\pm$ 0.2\\ 
ML32 & 1.3 &  190 $\pm$ 30 & 0.103 $\pm$ 0.015 & 7.7 $\pm$ 0.05 & 1.68  & 13 & 0.434 $\pm$ 0.013 & 55 $\pm$ 4 & 8.19 $\pm$ 0.06\\ 
\noalign{\smallskip} 
\hline
\end{tabular} 
\raggedright
\begin{tabular}{l}
\noalign{\smallskip} 
$^a$~Mean value taken from Paper~I. \\
$^b$~Inferred from the ratio between [SII]$\lambda$6716 and [SII]$\lambda$6731 \citep[e.g.,][]{1974agn..book.....O}.\\
$^c$~Half-light radius of the clump, measured from the FWHM of the H$\alpha$ flux distribution. \\
$^d$~Velocity dispersion of the clump, calculated from the H$\alpha$ line of the clump-integrated spectra.\\
$^e$~Dynamical mass of the clump from Eq.~(\ref{dispersion_mass}).
Errors do not include the uncertainty in the thermal motion subtraction. 
\\
$^f$~From \citet{2016ApJ...819..110S}.\\
$^g$~Assumed to be similar to the value in the other objects.\\ 
\end{tabular}
\label{tab:mean_data}
\end{table*}

The  velocity at each spatial pixel along the major axis of the galaxy was fitted 
with the universal rotation curve by \citet{2007MNRAS.378...41S}, i.e., 
 \begin{equation}
U(d)=U_0 + U_1 \frac{d-d_0}{\sqrt{\Delta^2+(d-d_0)^2}},
\label{universal_rc}
\end{equation}
where $U(d)$ stands for the observed  velocity at a distance \textit{d}, $U_0$  
represents the systemic velocity, i.e., the velocity at the dynamic center $d=d_0$, 
$\Delta$ gives a spatial scale for the central gradient of the rotation curve and, 
finally, $U_1$ provides the amplitude of the rotational velocity.  

The dynamical mass enclosed within a distance $d$, $M(d)$, follows from the balance between the 
centrifugal force and the gravitational pull, and it is given by,
\begin{equation}
\label{dispersion_mass_galaxy}
M(d)\,\sin^2i= (2.33 \times 10^5\,{\rm M}_\odot)\,(d - d_0)\,U^2(d),
\end{equation}
where $i$ stands for the inclination of the disk along the line-of-sight, 
and where distances are in kpc, and velocities in km\,s$^{-1}$.
The velocity $U$ in Eq.~(\ref{dispersion_mass_galaxy}) should be the
component along the line-of-sight of the 
circular velocity, $v_c\,\sin i$.  In the case of a purely stellar system, $U/\sin i$ and $v_c$
differ because the stellar orbits are not circular, with the difference given by the so-called  
asymmetric drift \citep[e.g.,][]{2001AJ....121..683H,2008gady.book.....B}. In our case, 
where we measure the velocity of the gas, they also differ because gas pressure gradients 
partly balance the gravitational force, and this hydro-dynamical force
has to be taken into account in the mechanical balance \citep[e.g.,][]{2010ApJ...721..547D}.
In a first approximation \citep[e.g.,][]{2010ApJ...721..547D,2016MNRAS.462.3628R},
\begin{equation}
v_c^2\simeq U^2/\sin^2 i+ 0.06\, W_U^2\,(d-d_0)/R_\star,
\label{eq:circular_veloc}
\end{equation}
where the coefficient that gives the correction has been inferred assuming the 
gas density to drop exponentially with the distance to the center ($|d-d_0|$),
with a length scale around three times the length scale of the stellar disks 
$R_\star$. This difference between the gaseous and the stellar disk is typical of XMPs 
\citep{2013A&A...558A..18F}.  Since $|d-d_0|\sim R_\star$,  Eq.~(\ref{eq:circular_veloc}) predicts 
that even for a large turbulent velocity ($W_U\simeq U/\sin i$), the difference between $v_c$ 
and $U/\sin i$ is small. Since we do not know the inclination angle of the galaxies, we cannot apply 
the correction for pressure gradients in Eq.~(\ref{eq:circular_veloc}). Thus, we 
employ $U$ in Eq.~(\ref{dispersion_mass_galaxy}), knowing that the resulting dynamical 
masses slightly underestimate the true masses in an amount given by $W_U/\sin i$ and 
Eq.~(\ref{eq:circular_veloc}).

Our targets have distinct star-forming regions (see Fig.~\ref{fig:all_images}),
also denoted here as star-forming clumps or starbursts.  
We refer to the brightest clump in each galaxy as the {\em main} star-forming region,
to be distinguished from other fainter clumps that appear in some galaxies.
They have been highlighted in Fig.~\ref{fig:all_images}.
The dynamical masses of the individual star-forming clumps are calculated from the velocity dispersion, 
assuming virial equilibrium, 
\begin{equation} 
\label{dispersion_mass}
M_{\rm dyn,turb}=(1.20\times 10^5\,{\rm M}_\odot)\, \ R_e \,{W_U}^2.
%\label{eq:dmasc}
\end{equation} 
The half-light radius, $R_e$, is calculated as in Paper~I, from the FWHM of a 1D Gaussian fitted to
the H$\alpha$ flux of the star-forming clump, which is subsequently corrected for 
seeing (Table~\ref{table_obs}). 
$W_U$ in Eq.~(\ref{dispersion_mass}) is calculated from the H$\alpha$ line resulting from 
integrating all the spectra along the spatial axis, from $-R_e$ to $+R_e$, with the interval centered at the clump. 
$R_e$ in Eq.~(\ref{dispersion_mass}) is given in kpc, and $W_U$ in km\,s$^{-1}$.
The use of $W_U$ to estimate dynamical masses implies excluding thermal motions in the 
virial equilibrium equation (Eq.~[\ref{eq:thermal}]). This approximation may be more or less appropriate depending  on the 
structure of the velocity field in the emitting gas. In practice, however, thermal motions
are always smaller than $W_U$, so that including them does not significantly
modify the masses estimated through Eq.~(\ref{dispersion_mass}).

Our long-slit spectra  only provide cuts across the star-forming regions. 
In order to obtain the total luminosity,  $\pounds_{\rm H\alpha}$, 
we assume the region to be circular, with a radius $R_e$. Then the flux obtained by integrating 
the observed spectra from $-R_e$ to $+R_e$, $F_{\rm H\alpha}$, is corrected according to 
the ratio between the 
observed area, $2\,R_e \times 1~{\rm arcsec}$, and the circular area, $\pi\,R_e^2$, which renders
\begin{equation}
\pounds_{\rm H\alpha}= 2\pi^2\, D^2\,{{R_e}\over{1~\rm arcsec}}\,F_{\rm H\alpha},
\end{equation}
with $D$ the distance to the source. We take $D$ from NED, as listed in Table~\ref{table_obs}.
SFRs are inferred from $\pounds_{\rm H\alpha}$ using the prescription by 
\citet{2012ARA&A..50..531K} given in Eq.~(\ref{kennicutt}).

%%%%%%%%%%%%%
\subsection{Characterization of the secondary components in the wings of H$\alpha$}\label{lobes} 

\begin{figure}
\includegraphics[width=\linewidth]{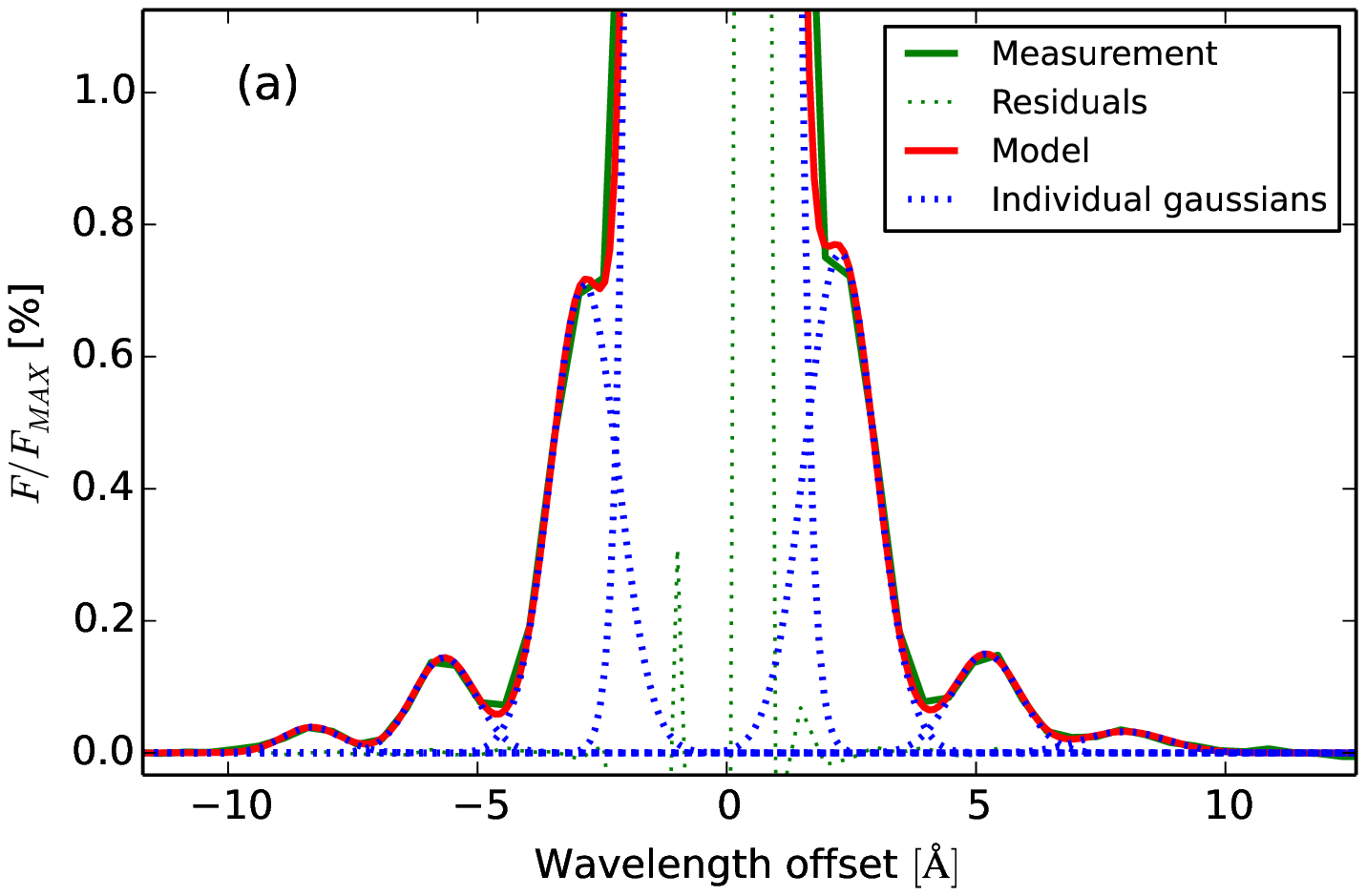}
\includegraphics[width=\linewidth]{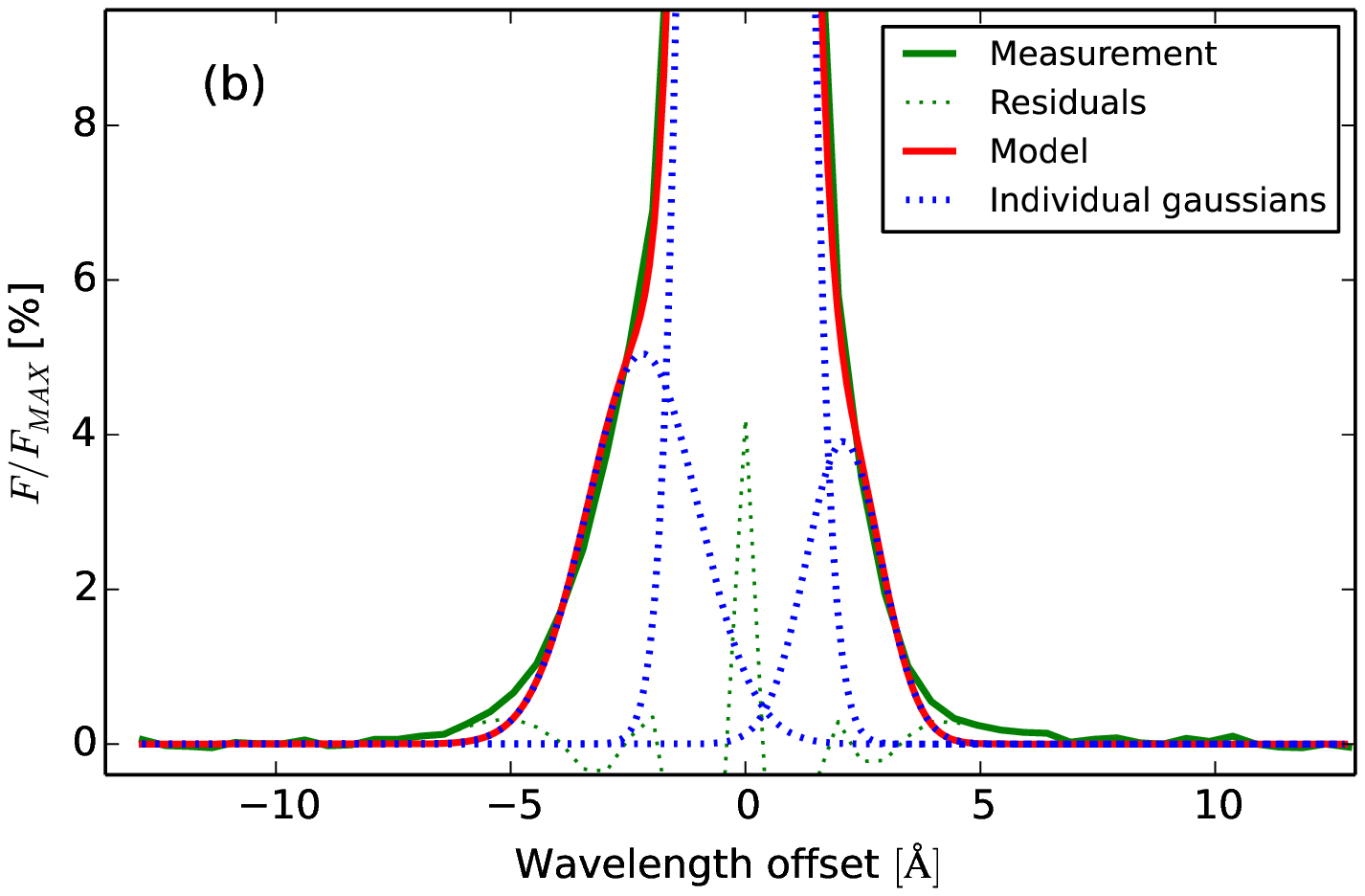}
\includegraphics[width=\linewidth]{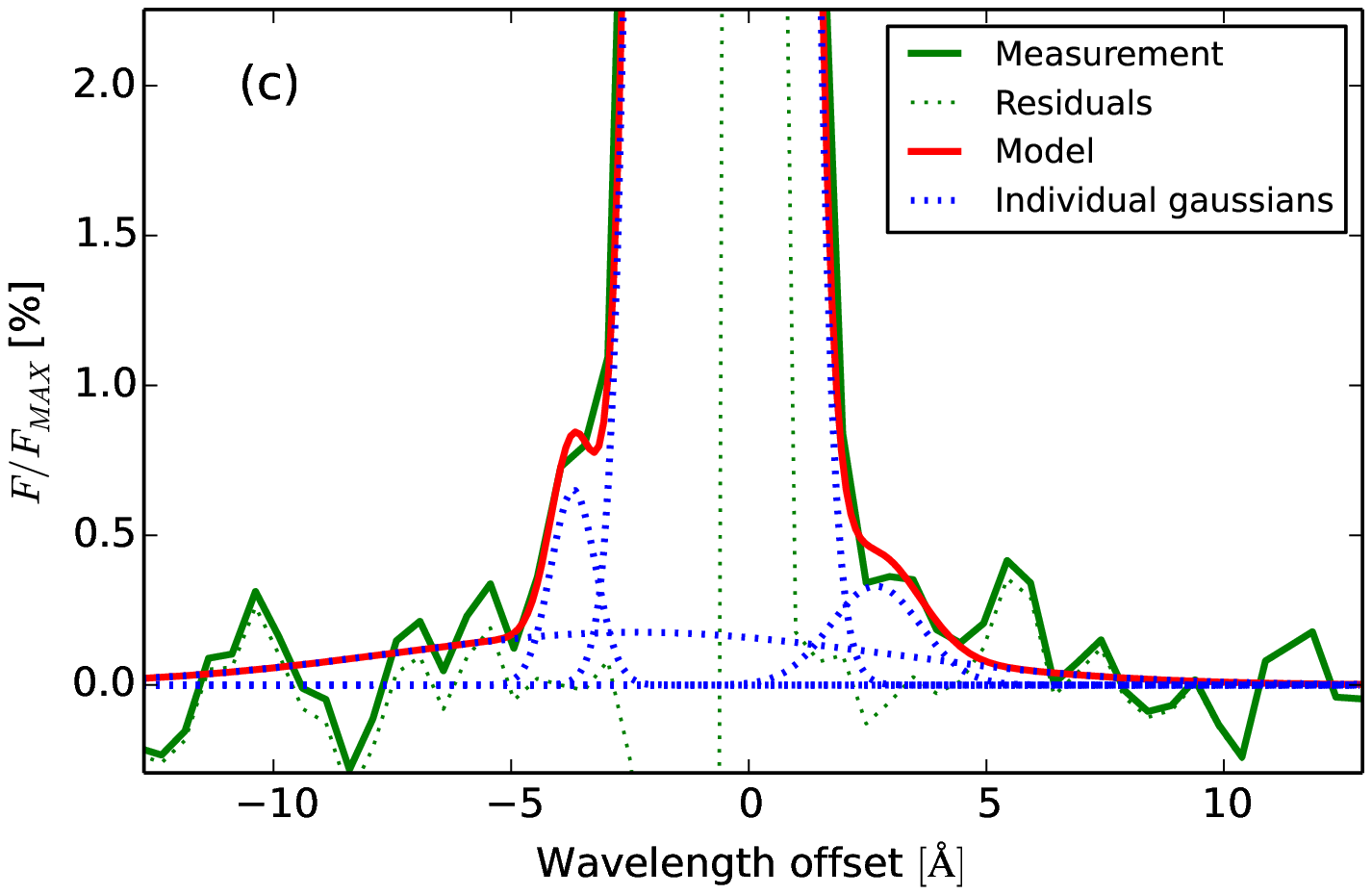}
\caption{
H$\alpha$ line  with the central component saturated 
to be able to distinguish the secondary components.
(a) H$\alpha$ line at the bright core of the galaxy UM461. 
The observed  spectrum (the green solid line) clearly shows seven components:
the central one plus three pairs.
Our fit reproduces the observation very well (the red solid line) 
as a superposition of Gaussian functions of various amplitudes, 
widths, and shifts (the blue dotted lines). The residuals 
of the fit are also included as the green dotted line.
(b) H$\alpha$ at the bright core of the galaxy SBS1, where
the components are not very well separated in wavelength.
(c) H$\alpha$ at a bright spot of N241, which 
portrays a case where the S/N is poor compared to
the previous cases.  
The wavelengths are referred to the centroid of the main component, and the fluxes 
are given in \%\ relative to the maximum flux of the H$\alpha$ profile. 
%normalized 
Each spectrum corresponds to a single spatial pixel.
}
\label{fig_lobes}
\end{figure}

When the S/N is sufficiently good, the observed H$\alpha$ profiles show faint broad emission, 
often in the form of multiple pairs of separate emission-line components (see Fig.~\ref{fig_lobes}).
The existence of pairs of components is
%Those lobes are 
particularly interesting from a physical standpoint, since they 
may reveal the feedback of star formation on the surrounding 
medium, or the presence of a black hole (BH; to be discussed 
in Sect.~\ref{sec:interpretation}).

In order to determine the properties of the secondary lobes (number, wavelength shifts, relative fluxes, etc.),
we performed a multi-Gaussian fit to H$\alpha$ using the Python package 
LMFIT \citep{newville_2014_11813}.  The fitting is not unique unless the free parameters 
are constrained, which we do by attempting to reproduce H$\alpha$ in the brightest pixel of the galaxy 
as a collection of discrete components, with widths similar to that of the central component.
The procedure is detailed below.
Several examples of the fit in various S/N conditions are given in Fig.~\ref{fig_lobes}. 
We initialize the fit in the spatial 
pixel with highest S/N in continuum, where the secondary components clearly stand out. 
This pixel corresponds to the brightest star-forming region of the galaxy (the bright blue knots in Fig.~\ref{fig:all_images}). 
Then the rest of the pixels are fitted, using as starting values the free parameters 
obtained in the fit to the adjacent pixel. 
The number of secondary components is set at the central brightest pixel, 
where the S/N needed to detect them is highest,  
and we fix this number for each galaxy. 
In order to follow each component across the spatial direction, we put two constraints on their spatial
variation,  namely, (1) the center does not deviate from the previous value
by more than twice the wavelength sampling ($\sim 1\,$\AA), and (2) the velocity dispersion is bound
between a minimum set by the spectral resolution, and twice the value in the previous pixel. 
These constraints implement the fact that neigh\-bouring pixels are not independent, since our
sampling grants at least two pixels per spatial resolution element. 
As a sanity check, we verified that the constraint on the center of the 
components does not affect the inferred velocities. The maximum wavelength 
displacement between two adjacent pixels for all secondary components in the clumps 
is approximately half the wavelength sampling.
Although the S/N decreases with the distance to the brightest pixel, 
we checked that each detected secondary component has signal at least 
twice the noise in continuum in all the spectra used to 
characterize the star-forming regions.

The uncertainties in each parameter can be derived from the co\-variance  matrix of the fit 
\citep[e.g.,][]{1986nras.book.....P}. However, when the fitted parameters are near the
values set by the constraints, the covariance matrix cannot be computed and the uncertainties 
provided by the Python routine turn out to be unreliable. Therefore, errors are 
estimated  running a MonteCarlo simulation. The best fit profile is contaminated with random 
noise having the observed S/N. This simulated observation is processed as a true observation,
and the exercise is repeated 80 times with independent realizations of the noise. 
The RMS fluctuations of the parameters thus derived are quoted as 1-sigma errors.
%%%%%%%%%%%%%%%%%%

\subsection{Mass loss rates}\label{mass_loss_rate}

In order to estimate the mass, the mass loss rate, and the kinetic energy corresponding 
to the secondary components observed in the wings of H$\alpha$ (Fig.~\ref{fig_lobes}),
we follow a procedure similar to that described in \citet{2015A&A...580A.102C}. 
It is detailed here both for completeness, and to tune the approach 
to our specific needs. 

We assume that the small emission-line components in the wings of the main H$\alpha$ profile
%emission line 
are produced by the recombination of H.
Then the luminosity in H$\alpha$, $L_{\rm H\alpha}$, is given by,
\begin{equation}
L_{\rm H\alpha}=\int\,f\,n_e\,n_p\,j_{\rm H\alpha}\,dV,
\label{eqone}
\end{equation}
where the integral extends to all the emitting volume,
and $n_e$, $n_p$, and $j_{\rm H\alpha}$ represent the number 
density of electrons and protons, and the emission coefficient,
respectively. The symbol $f$ stands for a local filling factor 
that accounts for a clumpy medium, so that only a fraction
$f$ is contributing to the emission. In a tenuous plasma, the 
emission coefficient is given by \citep{1974agn..book.....O,1999acfp.book.....L},
\begin{equation}
j_{\rm H\alpha}\simeq 3.56\times 10^{-25}\,t_4^{-1}~~{\rm erg\,cm^3\,s^{-1}},
\label{eqtwo}
\end{equation}
where $t_4$ is the temperature in units of $10^4\,$K. 
The mass of emitting gas turns out to be,
\begin{equation}
M_g=\int\,f\, m_{\rm H}\,n_{\rm H}\,X^{-1} \,dV,
\label{eqthree}
\end{equation}   
where $n_{\rm H}$ is number density of H (neutral and ionized), $m_{\rm H}$ stands for the 
atomic mass unit,  and  $X$ is the fraction of gas mass in H. 
If the gas is fully ionized\footnote{
If it is not, then $M_g$ and the other parameters inferred from this
mass refer to the mass of ionized gas.
}
($n_{\rm H}=n_p$), and the electron temperature and the H mass 
fraction are constant, then  Eqs.~(\ref{eqone}), (\ref{eqtwo}) and (\ref{eqthree}) render,
\begin{equation}
M_g={{m_{\rm H}\,L_{\rm H\alpha}}\over{X\,j_{\rm H\alpha}\,\langle n_e\rangle}_1},
\label{eqfour}
\end{equation}
with the mean electron density defined as, 
\begin{equation}
\langle n_e \rangle_{1}={{\int\,f\,n_p\,n_e\,dV}\over{\int\,f\,n_p\,dV}}.
\end{equation}
Using astronomical units, and assuming $X=0.75$ 
\citep[which corresponds to the solar composition; e.g.,][]{2009ARA&A..47..481A},
Eq.~(\ref{eqfour}) can be written as 
\begin{equation}
M_g=3.15\times 10^3\,{\rm M}_\odot~t_4\,
{{L_{\rm H\alpha}/10^{38}\,{\rm erg\,s^{-1}}}\over{\langle n_e\rangle_{1}/ 10^2\,{\rm cm^{-3}}}}.
\label{massgeq}
\end{equation}

\begin{figure}
\includegraphics[width=0.5\textwidth]{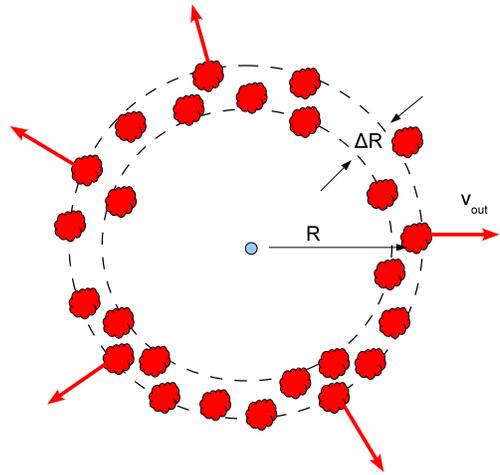}
\caption{
Schematic of the geometry used to estimate mass loss rates. 
The gas is represented in red as clumps which do not fill the 
full space. The schematic includes the characteristic length-scales, size
($R$) and width ($\Delta R$), and expansion velocity ($v_{\rm out}$). 
The blue dot points out the center of expansion.
}
\label{cartoon}
\end{figure}

Assume that the motions associated with the weak emission-line components in the wings of the main 
profile trace some sort of expansion. The mass loss rate, $\dot{M}_g$,
is defined as the total mass carried away per unit time by these motions. It can
be computed as the flux of mass across a closed surface around the center of expansion,
\begin{equation}
\dot{M}_g=\int\, f\,\rho_g\, {\bf v\cdot d\Sigma},
\end{equation}
where $\rho_g$ is the gas density at the surface and ${\bf v}$ represents the
velocity of the flow. Assuming that motions are radial, and that
the density and filling factor are constant at the surface,
the previous integral can be simplified considering a spherical surface of radius $R$, 
\begin{equation}
\dot{M}_g=4\pi\,R^2\,f\,\rho_g v_{\rm out},
\end{equation}
with the radial speed $v_{\rm out}=|{\bf v}|$. If the moving mass forms a 
shell of width $\Delta R$ (Fig.~\ref{cartoon}), 
then $M_g=4\pi\,R^2\,f\rho_g\,\Delta R$, so that the mass loss rate turns out to be  
%If the gas is moving out at a radial speed $v_{\rm out}$, then the mass-loss rate is 
\begin{equation}
\dot{M}_g\simeq M_g\,v_{\rm out}\,\Delta R^{-1}.
\label{massloss}
\end{equation}
If the moving mass is distributed on a sphere, then $M_g=4\pi\,R^3\,f\rho_g/3$, and the 
mass loss rate is still given by  Eq.~(\ref{massloss}) provided that $\Delta R/R\simeq 1/3$. 
In other  words, by using $\Delta R/R\simeq 0.3$ in Eq.~(\ref{massloss}) one approximately considers 
the full range of geometries between shells and spheres (for further discussion on the assumptions made to estimate $\dot{M}_g$ see, e.g.,
\citeauthor{2012MNRAS.425L..66M}~\citeyear{2012MNRAS.425L..66M};
\citeauthor{2014MNRAS.442.3013K}~\citeyear{2014MNRAS.442.3013K}; or
\citeauthor{2015A&A...580A.102C}~\citeyear{2015A&A...580A.102C}).

Once the expanding shell geometry has been 
assumed, the radius of the shell is set by the gas mass and the electron density, since 
Eq.~(\ref{eqthree}) can be written as, 
\begin{equation}
M_g\simeq {{8\pi\,m_H}\over{1+X}}\,\langle n_e\rangle_2\,\langle f\rangle_3\,R^2\,\Delta R,
\label{eqfive}
\end{equation} 
with the mean values given by 
\begin{displaymath}
\langle n_e \rangle_{2}={{\int\,f\,n_e\,dV}\over{\int\,f\,dV}},
\end{displaymath}
\begin{displaymath}
\langle f \rangle_{3}={{\int\,f\,dV}\over{\int\,dV}}.
\end{displaymath}
Equation~(\ref{eqfive}) assumes the gas to be composed of fully ionized H and He, so that 
\begin{equation}
n_e\simeq n_{\rm H}+2 n_{\rm He}\simeq n_{\rm H} {{1+X}\over{2X}},
\end{equation}
which is a reasonable estimate considering that we are dealing with metal-poor gas, such that the
metals do not contribute to the electron density. Equation~(\ref{eqfive}) provides
a way to estimate the size of the emitting region in terms of the electron density, the 
filling factor, and the relative width of the shell, explicitly,
\begin{equation}
\Big[{{R}\over{14.7~{\rm pc}}}\Big]^3=\Big[
{{M_g}\over{10^4\,{\rm M}_\odot}}\Big]
\Big[{{\langle n_e\rangle_2}\over{10^2\,{\rm cm}^{-3}}}
{{\langle f \rangle_3}\over{0.3}}
{{\Delta R/R}\over{0.3}}
\Big]^{-1}.
\label{size1}
\end{equation}
The expansion velocity and the radius allow us to estimate the age of the expanding shell,
or the time-lag from blowout,
\begin{equation}
Age=R/v_{\rm out},
\label{eq:age_shell}
\end{equation}
where $v_{\rm out}$ is assumed to be constant in time.
Since the expansion decelerates, $Age$ in Eq.~(\ref{eq:age_shell}) represents 
an upper limit to the true age.

A convenient (and intuitive) way of expressing the mass loss rate in Eq.~(\ref{massloss})
is re-writing the expression in terms of the Star Formation Rate (SFR)
estimated using the prescription by 
\citet{2012ARA&A..50..531K}\footnote{This SFR is 30\% smaller than 
the classical value in \citet[][]{1998ARA&A..36..189K}.}, 
\begin{equation}
{{\rm SFR}\over{\rm 1 M_\odot yr^{-1}}}\simeq {{\pounds_{\rm H\alpha}}\over{1.86\times 10^{41}\,{\rm erg\,s^{-1}}}},
\label{kennicutt}
\end{equation}
which depends only on the total H$\alpha$ flux of the region, $\pounds_{\rm H\alpha}$.
We parameterize the H$\alpha$ flux in a weak emission-line component, 
$L_{\rm H\alpha}$, in terms of total H$\alpha$ flux,
\begin{equation}
L_{\rm H\alpha}=\varepsilon \pounds_{\rm H\alpha},
\label{epsilon}
\end{equation} 
with $\varepsilon$ the scaling factor
inferred from the fits described in Sect.~\ref{lobes}.
We do not correct $\pounds_{\rm H\alpha}$ for extinction in Eq.~(\ref{kennicutt})
because the photometric calibration needed to infer it from the Balmer decrement 
is uncertain (Sect.~\ref{data_reduction}), 
because the main H$\alpha$ lobe and the weak emission components do not 
necessarily have the same extinction, and because the expected extinction is very 
low \citep[reddening coefficient around 0.1;][]{2016ApJ...819..110S}.
Using Eqs.~(\ref{kennicutt}) and (\ref{epsilon}), Eq.~(\ref{massloss})
can be expressed as 
\begin{equation}
{{\dot{M}_g}\over{\rm SFR}}\simeq 0.20 %0.041%\times10^{-2}
{{\varepsilon}\over{10^{-3}}}
{{v_{\rm out}}\over{10^2\,{\rm km\,s^{-1}}}}\,
t_4\,
\Big[{{\Delta R}\over{3\,{\rm pc}}}\,
{{\langle n_e\rangle_1}\over{10^2\,{\rm cm^{-3}}}}\Big]^{-1}.
\label{massloss2}
\end{equation}

Another important parameter derived from the gas mass and velocity is 
the kinetic energy involved in the outflow,
\begin{equation}
E_k={{1}\over{2}}M_g\,v_{\rm out}^2\simeq 10^{51}\,{\rm erg}\,{{M_g}\over{10^4\,{\rm M_\odot}}}\,
\Big[{{v_{\rm out}}\over{10^2\,{\rm km\,s^{-1}}}}\Big]^2.
\label{eq:kinetic}
\end{equation} 
The equation has been written in units of  $10^{51}{\rm erg}$, which is the typical kinetic energy 
released in a core-collapse supernova  \citep[SN; e.g.,][]{1999ApJS..123....3L,2009ApJ...703.2205K},
corresponding to the explosion of a massive star.

This energy range is also within reach of accreting BH of intermediate 
mass, like the direct collapse $10^5~{\rm M}_\odot$ BH seeds needed to explain the 
existence of super-massive BHs in the early Universe 
\citep[e.g.,][]{2010A&ARv..18..279V,2014MNRAS.443.2410F,2016MNRAS.459.1432P}. 
Assuming that the BH releases kinetic energy at a fraction of the Eddington limit, $\psi$, 
and that the accretion event has a duration, $\Upsilon$, then
\begin{equation}
E_{BH}\simeq 10^{51}\,{\rm erg}\, {{\psi}\over{0.1}}\, {{\Upsilon}\over{25~{\rm yr}}}\,{{M_{BH}}\over{10^5\,{\rm M}_\odot}},
\label{eq:bh_mass2}
\end{equation}
with $M_{BH}$ the BH mass \citep[e.g.,][]{2002apa..book.....F}.

%%%%%%%%%%%%%%%%

\subsection{Mass of the central object}\label{sec:mass_bh}

In principle, the observed faint emission-line pairs in the wings of H$\alpha$ might also be generated in a rotating disk 
around a massive object \citep[e.g.,][]{1964AJ.....69..521E}. Then the approximation used to estimate the radius 
of the shell in Eq.~(\ref{size1}) still holds, except that 
\begin{equation}
\Big[{{R}\over{27.6~{\rm pc}}}\Big]^3=\Big[
{{M_g}\over{10^4\,{\rm M}_\odot}}\Big]
\Big[{{\langle n_e\rangle_2}\over{10^2\,{\rm cm}^{-3}}}
{{\langle f \rangle_3}\over{0.3}}
{{\Delta R/R}\over{0.3}} {{\Delta z/R}\over{0.3}}
\Big]^{-1},
\label{size2}
\end{equation}
where $\Delta R$ represents the difference between the inner and outer 
radii, and $\Delta z$ stands for the thickness of the disk.
This size, together with the circular velocity, $v_{c}$, directly inferred from the spectra,
allow us to estimate the mass of the central object, $M_\bullet$, 
from the balance between gravity and centrifugal force,
\begin{equation}
{M_\bullet}\simeq {7.0\times 10^7\,{\rm M}_\odot}\,{{R}\over{30\,\rm pc}}\,\Big({{v_c}\over{10^2\,{\rm km\,s^{-1}}}}\Big)^2.
\label{eq:mass_bh}
\end{equation}

%%%%%%%%%%%%%%%%

\section{Rotation, turbulent motions, and chemical properties}\label{sect:rot_turb}

\begin{figure*}
\centering 
\includegraphics[width=\linewidth]{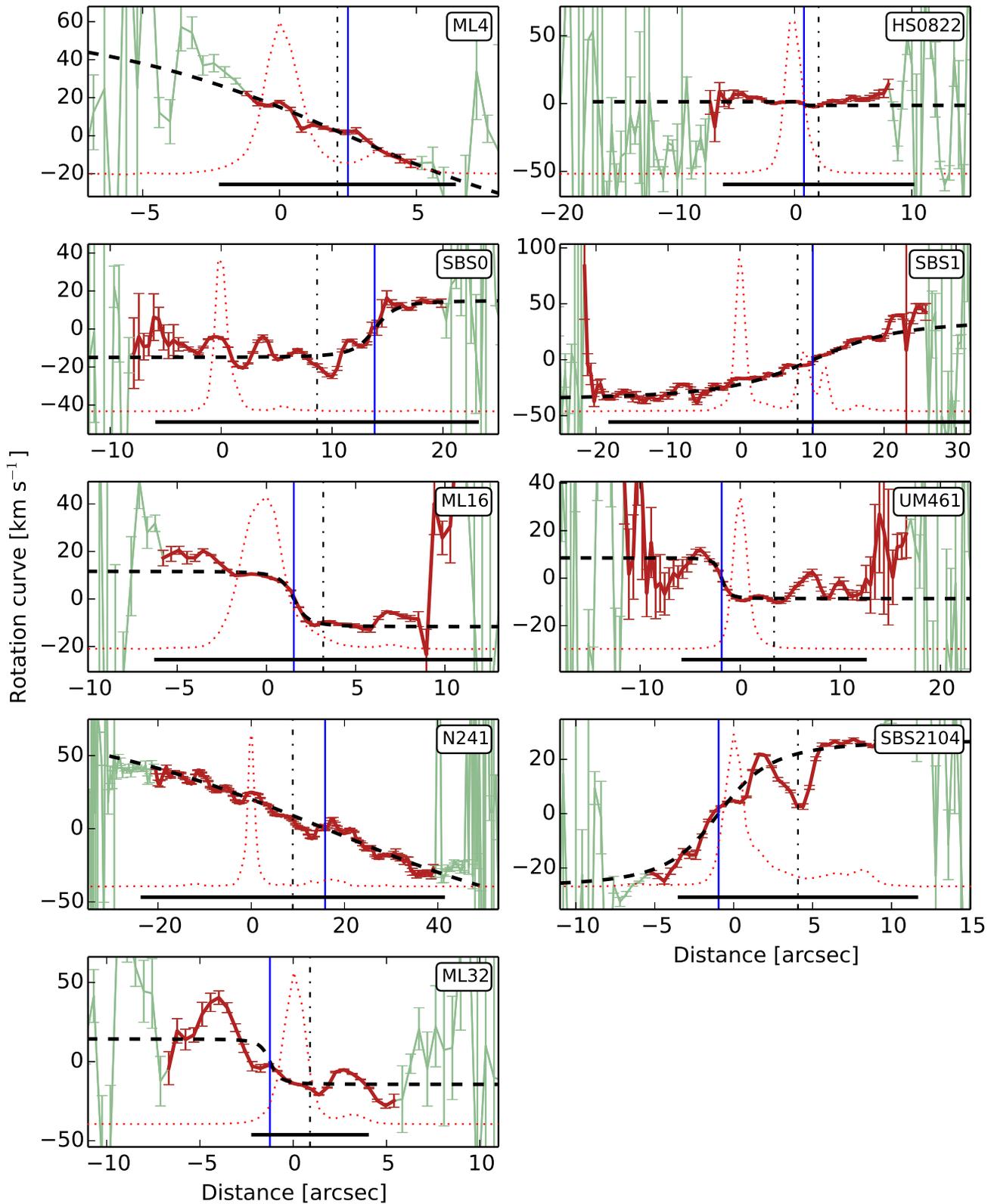}
\caption{Rotation curves for the XMPs, i.e., velocity versus position along the 
slit. The velocities were derived using a Gaussian fit to the H$\alpha$ line profile. 
The actual observed points show error bars joined by a continuous line, which is either 
green or red, depending on whether the S/N of the spectra in the continuum is
smaller (green) or larger (red) than one. 
Only the points in red have been used to fit the universal rotation curve, shown as
the black dashed line, with the vertical blue line 
identifying the position of the dynamical center inferred from the fit.
The zero of the velocity scale is set by the fit at the dynamical center.
The red dotted line represents the H$\alpha$ flux scaled from zero to the maximum value.
Distances are referred to the point with largest H$\alpha$ flux.
The horizontal solid black line gives the galaxy diameter, inferred from the 
25~mag arcsec$^{-2}$ isophote, and provided by NED. It is centered in the 
photometric midpoint marked by the vertical dotted-dashed line. 
}
\label{fig:rot_curve}
\end{figure*}

\begin{figure*}
\includegraphics[width=\linewidth]{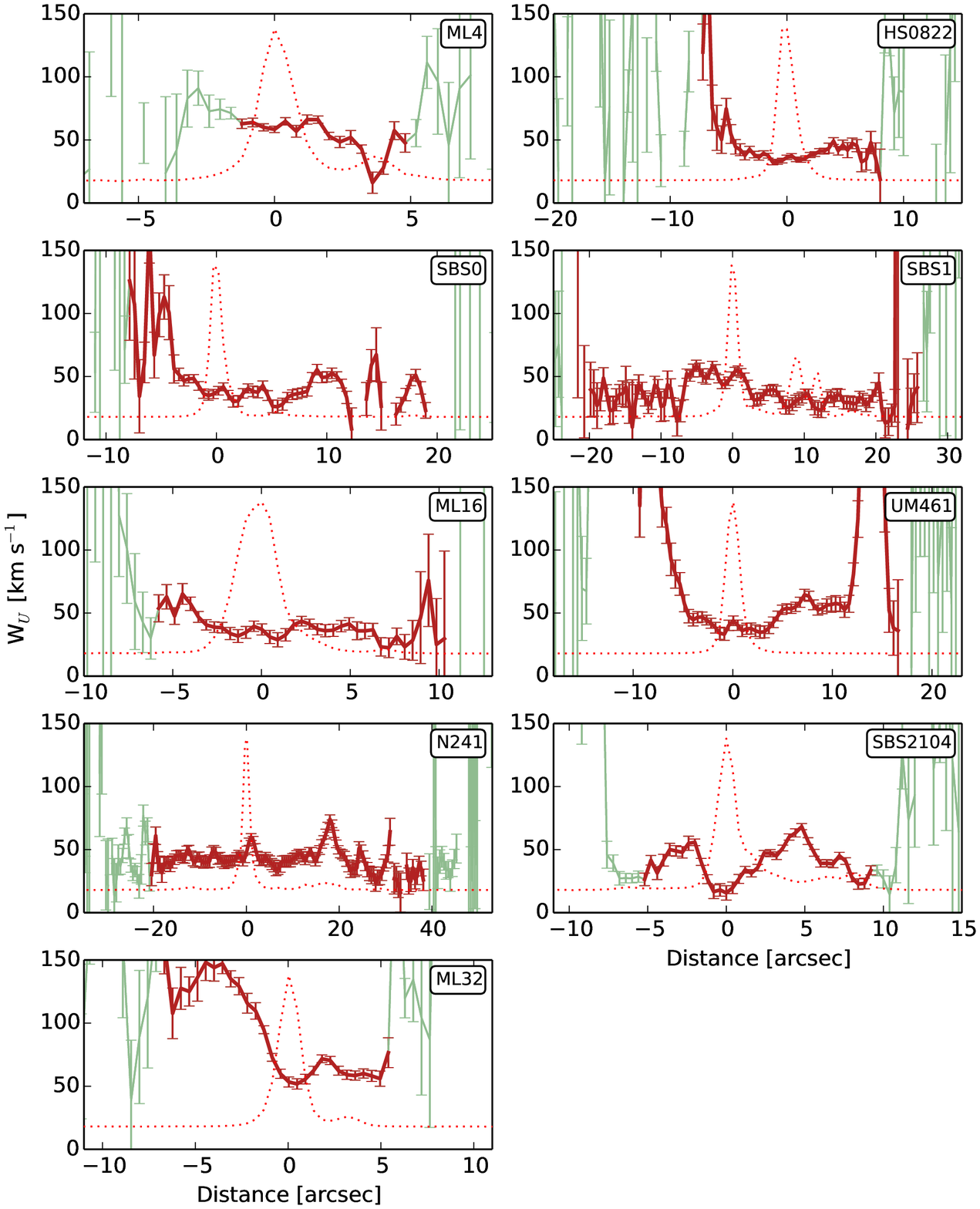}
\caption{Similar to Fig.~\ref{fig:rot_curve} with the FWHM velocity dispersion.
The range in abscissas is the same as in Fig.~\ref{fig:rot_curve}. 
}
\label{fig:dispersion}
\end{figure*}

Figure \ref{fig:rot_curve} shows the rotation curve of all the galaxies analyzed in this 
study.  It represents the velocities inferred from the Gaussian fit, but the results are similar 
when using centroids (Sect.~\ref{main}). 
Most galaxies seem to rotate, although HS0822 and SBS0 do not.   
The rotation curve of UM461 may suggest some rotation, but it  may also be 
compatible with the absence of rotation. The difficulty in detecting rotation in UM461 
could be due to projection effects; UM461 appears face-on in the SDSS image, with the galaxy showing a rounded faint envelope enclosing its two bright 
knots (Fig.~\ref{fig:all_images}).
All rotation curves present small-scale velocity irregularities, often with an amplitude comparable 
to the  velocity gradient across the whole galaxy.  The rotation of the galaxies ML4, SBS1, N241 is clear,
whereas other galaxies are more affected by these small-scale irregularities.   
Galaxies SBS2104 and ML32 could host counter-rotating components, like those found in 
tadpole galaxies by \citet{2013ApJ...767...74S},  although, in this case, the center of the 
counter-rotating structure does not match the main star-forming region. 
Figure~\ref{fig:dispersion} shows the FWHM velocity dispersion inferred from H$\alpha$.
Typical values at the position of the main starburst are between 20 and 60~km\,s$^{-1}$ 
(Table~\ref{tab:mean_data}). We note that the regions with possible counter-rotating 
components also show an increase
in the velocity  dispersion: c.f., Figs.~\ref{fig:rot_curve} and \ref{fig:dispersion} for SBS2104 ($d\simeq 4$~arcsec)
and ML32 ($d\simeq 2$~arcsec).

We fitted the universal rotation curve by \citet{2007MNRAS.378...41S}, given in Eq.~(\ref{universal_rc}), 
to the observed velocities. The four free parameters of the non-linear fit
($U_0$,$U_1$, $d_0$, and $\Delta$)
were derived using the 
Python package LMFIT \citep{newville_2014_11813}. The fits are shown as black dashed lines in Fig.~\ref{fig:rot_curve}.
The fits are not particularly good, but they allow us to have an idea of the amplitude of the rotation
curve and the dynamical center of the galaxies. These properties are summarized in Table~\ref{table_rotation}.   
Even though the position of the dynamical center is very uncertain (see Fig.~\ref{fig:rot_curve}),
they do not seem to overlap with the position of the main star-forming region.

Independently of the properties of the rotation, we do find 
a recurrent behavior for the velocity and velocity dispersion  at the star-forming region. 
The velocity tends to be constant within the spatial extent of the clump. This is clear 
in, e.g., SBS2104 and ML16 (see the velocity curves in Fig.~\ref{fig:rot_curve} at the point where 
the H$\alpha$ emission peaks). If the spatial resolution of the observations were insufficient to 
resolve the individual clumps, then one would expect that the velocity, and all the 
other properties inferred from the shape of the H$\alpha$ line, would be constant
within the spatially unresolved clump. However, insufficient spatial
resolution is not responsible for the observed behavior for 
two reasons: (1) many clumps are well resolved; e.g., ML16 has a FWHM of 
3~arcsec (Fig.~\ref{fig:rot_curve}) when the seeing was only 1~arcsec FWHM
(Table~\ref{table_obs}), and (2) the velocity dispersion is not constant 
within the starburst clump (see Fig.~\ref{fig:dispersion}).

\begin{figure}
\centering 
\includegraphics[width=\linewidth]{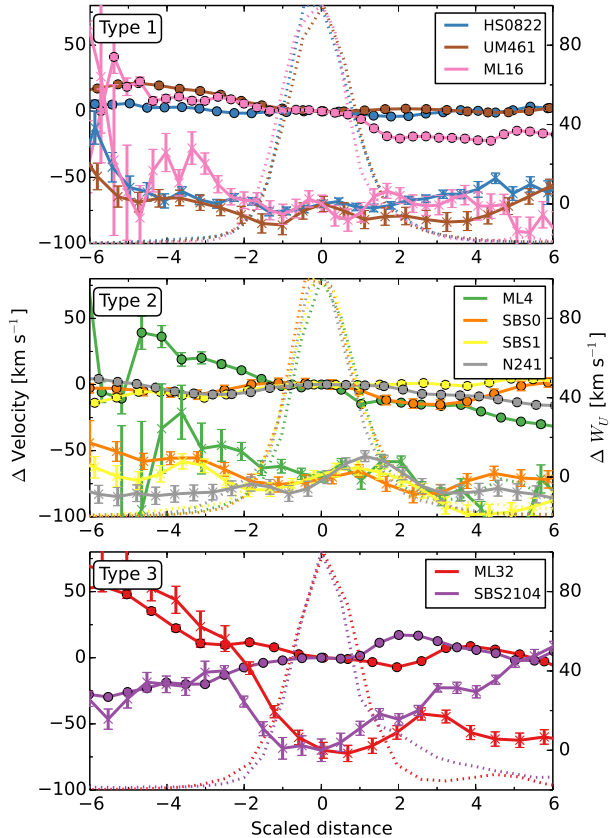}
\caption{Variation of the velocity (bullet symbols) and velocity dispersion (asterisks)
across the star-forming regions of the XMPs. Distances, in the abscissa axis, are referenced to the 
position of the star-forming region and normalized to its size. 
For this reason, the variation of the H$\alpha$ flux across the region, shown 
as dotted lines, is always centered at distance zero, and always has 
the same width. The velocity turns out to be constant within the star-forming clump, 
whereas the velocity dispersion follows
three distinct patterns, 
labelled as Type~1 (shown in the top panel), Type~2 (middle panel), and Type~3 (bottom panel). 
In Type~1, the dispersion presents a local maximum at the center of the starburst. In Type~2, however, 
the dispersion increases across the region towards the position that is closest to the 
galaxy photometric center. 
The Type~3 pattern presents a minimum at the center of the star-forming region.
Different colors refer to different galaxies, as indicated by the insets. 
The monikers used for the XMPs are in Table~\ref{table_obs}.
%\red{Amanda: check that the color code is the same as that used throughout.}
%
The average velocity and velocity dispersion in each star-forming clump have been 
subtracted from the plots, so that the velocity and velocity
dispersion are zero at the clump.
}
\label{fig:disp}
\end{figure}

These two properties, i.e., constant velocity with varying velocity dispersion,
are clearer in Fig.~\ref{fig:disp}.  
It shows the H$\alpha$ flux across all the clumps of the individual
galaxies (the dotted lines in all the panels). 
They have been re-scaled in abscissa and ordinate  so that all of the H$\alpha$ flux profiles look 
approximately the same in this representation (we normalize each profile to its maximum flux and to its size). 
The figure also includes the rotation curve around the star-forming region (bullet 
symbols and solid lines), once the average velocity in the region has 
been removed. Note that all rotation curves tend to flatten at the location 
of the star-forming clump. The velocity is uniform, as if the star-forming region
were a kinematically distinct entity. Note also how the velocity dispersion (shown as asterisks
joined by solid lines in Fig.~\ref{fig:disp}) varies across the clump.
Three patterns for the variation of the velocity dispersion can be distinguished, 
namely, the velocity dispersion has a local maximum coinciding with the center of the star-forming clump
(Fig.~\ref{fig:disp}, top panel; Type 1), the velocity dispersion 
increases across the clump (Fig.~\ref{fig:disp}, middle panel; Type 2),
and the velocity dispersion presents a minimum at the center of the clump (Fig.~\ref{fig:disp}, bottom panel; Type 3). 

The first case may be qualitatively understood if the star-forming region is undergoing
a global expansion. The velocity dispersion peaks in the central part, where the Doppler 
signal of the approaching and receding parts is largest. In this instance, the  
velocity remains constant, tracing the rotational velocity at the origin of 
expansion\footnote{Even if the original region had some rotation, it would have been washed out 
by the expansion, due to angular momentum conservation.}. 
The difference between the velocity dispersion at the center and at the sides of the 
starburst can be used to estimate the expansion velocity of the region. If the difference 
is of the order of 20\,\%, and 
$W_U$ is around 40~km\,s$^{-1}$ (see HS0822, UM461, and ML16 in Table~\ref{tab:mean_data}), 
then an expansion able to create the observed excess of central velocity dispersion 
has to be of the order of  13~km\,s$^{-1}$. We have assumed that
the turbulent and expansion velocity add up quadratically in their 
contribution to $W_U$. The mass loss rate resulting from the possible global expansion
is worked out in Sect.~\ref{sec:mydiscussion}.
%The first case is similar to the (stellar) velocity dispersion in globular
%clusters \citep[GC; e.g.,][]{2014MNRAS.445.4446K}, and in the case of GCs, 
%the decrease is due to \red{I do not know ... but may be the same explanation in this case}.

In the second case (Type 2; Fig.~\ref{fig:disp}), the dispersion always increases towards positive
distances which, by construction, point to the photometric center of the galaxies.
In other words, it is largest in the part of the star-forming clump that is closest 
to the galaxy center. 
As we discussed above, this is not an observational bias due to insufficient spatial resolution.
Although we can only speculate at this 
point, the result is very suggestive.  The excess velocity dispersion in the center-side 
of Types~2 may indicate an intensification of the turbulence of the gas in 
that particular part of the starburst.  Such an increase may indicate the 
collision of the starburst gas with the ISM of the host galaxy,  
as if the star-forming regions were inspiraling towards the galaxy center.
This migration to the galaxy center is expected from tidal forces 
acting upon massive gas clumps
\citep[see, e.g.,][]{2008ApJ...688...67E,2016MNRAS.457.2605C,2016arXiv160401698H}; the large starbursts in our XMPs may be going through such a process at this
moment. 

The Type 3 pattern presents a minimum velocity dispersion, 
coinciding with the starburst (Fig.~\ref{fig:disp}); it may reflect the past expansion of the region. As an adiabatic contraction increases
the turbulence in a contracting medium \citep[e.g.,][]{2015ApJ...804...44M},
an expansion leads to adiabatic cooling and to a drop of the turbulence 
\citep[e.g.,][]{2012ApJ...750L..31R}. A past expansion phase may have produced the
observed decrease in velocity dispersion, even if the phase is already over
and it does not appear in the form of a Type~1 pattern.
The same kind of local minima in velocity dispersion has been observed in 
young star cluster complexes of nearby galaxies 
\citep[e.g.,][]{2006A&A...445..471B}.

We have calculated the dynamical mass that accounts for the velocity dispersion 
if the clump was in virial equilibrium, as described by Eq.~(\ref{dispersion_mass}). 
The results are in  Table~\ref{tab:mean_data}. They are similar to the 
total stellar mass of the galaxy and thus, very large. 
They are also comparable to the dynamical mass of the galaxy inferred from the rotation curve 
(Table~\ref{table_rotation}), even though these masses are lower limits because of the (unknown) 
inclination of the galaxies, and because of the existence of pressure gradients 
(see the discussion in Sect.~\ref{explaining_rules}).

The fact that the star-forming regions have well defined kinematical properties, 
%different from the rest of the disk, 
suggests that they are %kinematically 
distinct 
entities within the host galaxy.
\begin{table}[!h] 
\caption{Properties of the rotation curve}
\centering 
\begin{tabular}{lcccc} 
\hline 
\noalign{\smallskip} 
ID   & $U_1\,^a$ & $d_0\,^b$  & $\log(M\,\sin^2{i})\,^c$ & $|d_\mathrm{l}-d_0|\,^c$ \\ 
  & [km\,s$^{-1}$] &  [arcsec] &  [M$_\odot$] &  [arcsec] \\ 
\noalign{\smallskip} 
\hline 
\noalign{\smallskip} 
ML4  & -70 $\pm$ 200  & 2 $\pm$ 2 & 8.4 & 3.7 \\ 
HS0822  & -1.3 $\pm$ 0.2 & 0.8 $\pm$ 80 & 5.2 & 7.5 \\ 
SBS0 & 15 $\pm$ 1.2  & 13.8 $\pm$ 0.5 & 8.1 & 21.7 \\ 
SBS1 & 36 $\pm$ 4 & 10.1 $\pm$ 0.08 &  9.0 & 31.7 \\ 
ML16  & -11.7 $\pm$ 0.7 &  1.53 $\pm$ 0.09 &  7.4 & 7.6 \\ 
UM461 & -8.6 $\pm$ 0.7 &  -1.85 $\pm$ 0.08 &  7.2 & 13.4 \\ 
N241 & -90 $\pm$ 50 &  16 $\pm$ 10 & 9.0 & 40.8 \\ 
SBS2104 & 27 $\pm$ 8 &  -1 $\pm$ 0.9 & 8.2 & 10.2 \\ 
ML32 & -14 $\pm$ 7 & -1.2 $\pm$ 0.4 & 8.3 & 6.6\\ 
\noalign{\smallskip} 
\hline 
\end{tabular} 
\begin{tabular}{l}
\noalign{\smallskip} 
$^{a}$~Amplitude resulting from fitting Eq.~(\ref{universal_rc}) to the observed\\ ~~~rotation curve.\\
$^{b}$~Dynamical center resulting from fitting Eq.~(\ref{universal_rc}) to the\\ ~~~observed rotation curve.\\
$^c$~Dynamical mass of the galaxy. Based on Eq.~(\ref{dispersion_mass_galaxy}), considering \\ 
~~~up to the largest distance from $d_0$ with sensible velocity 
\\~~~measurements, d$_\mathrm{l}$.\\
%\referee{$^d$~The last radius used to calculate the dynamical mass.}\\ ~~~ \referee{ Measured from the dynamical center.}\\
\end{tabular}
\label{table_rotation}
\end{table}

\subsection{Chemical properties}
%%%%
\begin{figure*}
\includegraphics[width=\linewidth]{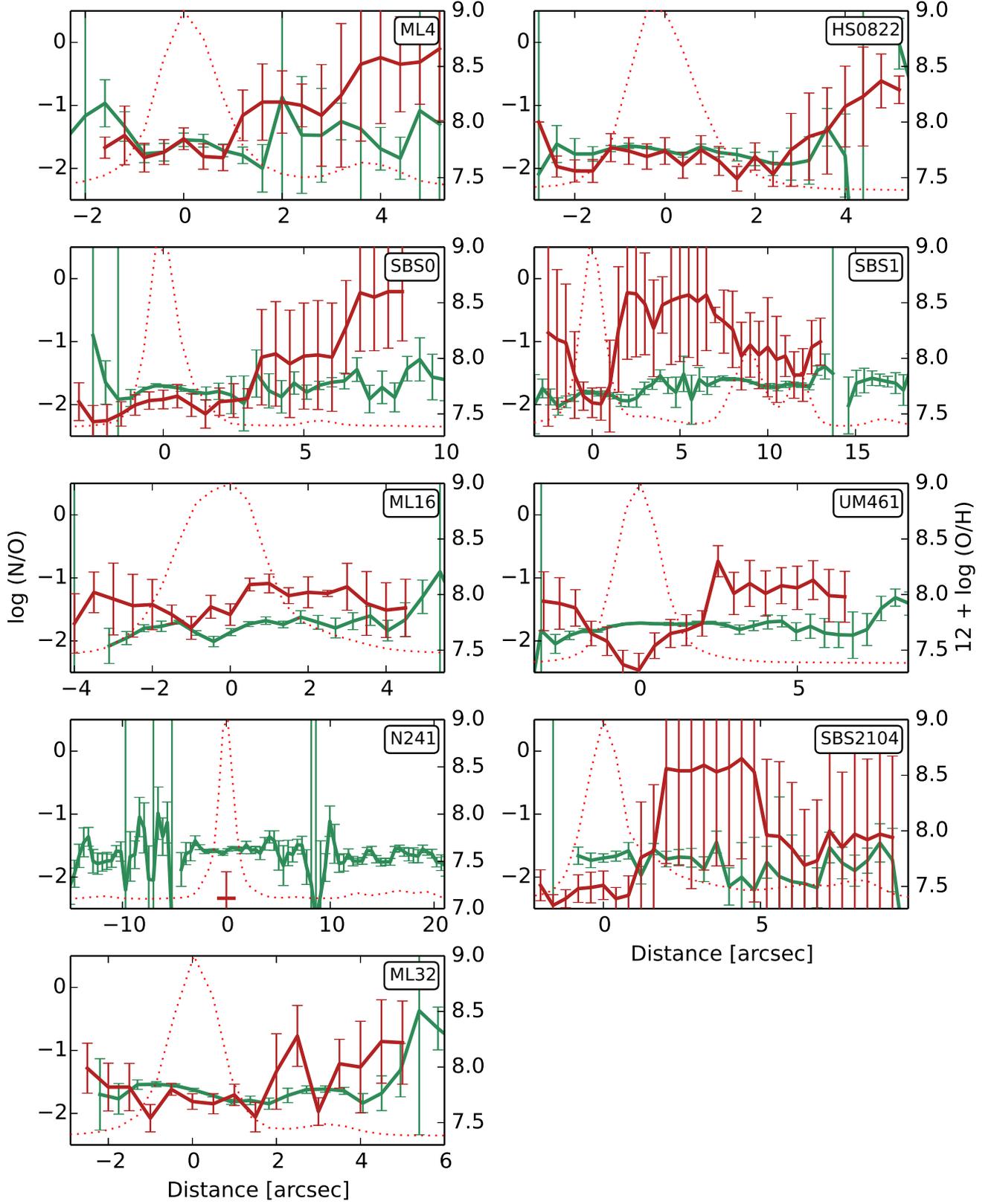}
\caption{
Variation of the metallicity ($12+\log[{\rm O/H}]$), N/O ratio, and SFR along the major axis of the XMPs. The 
red solid line is the metallicity of the galaxy from Paper~I, 
except for N241, where a mean value for the clump was obtained from \citet{2016ApJ...819..110S}. 
The green solid line is the variation of the N/O ratio. 
%The black solid line at $\log({\rm N/O}) = 1.5$ is a common value in low metallicity galaxies.  
The red dotted line represents the H$\alpha$ flux scaled to the maximum value at the clump, which is 
a proxy for the SFR.
}
\label{fig:oh_no}
\end{figure*}

A separate, but important, result from the analysis of the spectra deals with the metallicity. 
Most of the XMPs that we study are characterized by having a drop in gas-phase
metallicity associated with 
the main starburst (Paper~I). We searched for possible relationships between such metallicity
inhomogeneities and the kinematic properties described above but found no correlation. 
What we found, however, is a notable lack of correlation between the metallicity (as traced by O/H) 
and the ratio N/O. It appears in Fig.~\ref{fig:oh_no}, where the variation of the metallicity, SFR and N/O ratio is displayed.
The N/O ratio was estimated from [NII]$\lambda$5583 and [SII]$\lambda$6716,6731
following the calibration by \citet{2009MNRAS.398..949P}, which is almost independent 
of dust reddening (it was not estimated in Paper~I because of the difficulty in separating [NII] from H$\alpha$, given 
the limited wavelength resolution of that spectra). Figure~\ref{fig:oh_no} shows that the N/O ratio remains
fairly constant along the galaxy, even at the position of the main starbursts, where the H$\alpha$ flux
peaks and the metallicity drops. The lack of correlation between the ratios N/O and O/H represents the general behavior.
This observational fact is consistent with the gas accretion scenario. If the accretion
of metal-poor gas is triggering the observed starburst, then the fresh gas would locally reduce the O/H ratio in the 
star-forming region relative to the rest of the galaxy. However, mixing with external gas cannot modify the 
pre-existing ratio between metals, leaving the N/O ratio unchanged. This argument has been put forward before 
to support the gas accretion scenario \citep[][]{2010ApJ...715L.128A,2012ApJ...749..185A,2014A&ARv..22...71S}.
The value for $\log({\rm N/O})$ that we infer, around $-1.5$, is also consistent with the N/O ratio found in 
metal-poor $\alpha-$enhanced stars in the solar neighborhood 
\citep[e.g.,][]{2004A&A...421..649I,2005A&A...430..655S}. 
Since the star formation is enhanced where the O/H ratio drops, the lack of correlation between the ratios 
N/O and O/H reveals a lack of correlation between N/O and the SFR. This property seems to be common among  
star-forming galaxies at all redshifts \citep[e.g.,][]{2013A&A...549A..25P}, and supports the 
metal-poor nature of the gas that forms stars.

%%%%%%%%%%%%%%%%%%%%%%%%%%%%%%%
\section{Properties of the multiple components}\label{multiple_comp}

\subsection{Expanding shell interpretation}\label{sec:mult1}

\begin{figure}
\includegraphics[width=0.5\textwidth]{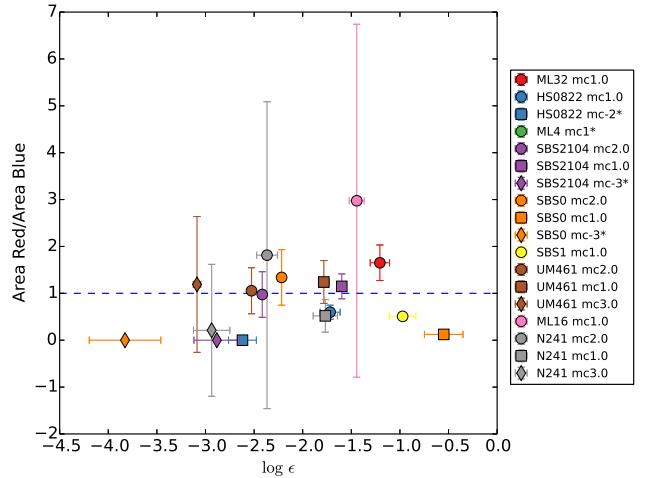}\\%jpg}\\
\caption{Area ratio between the blue and red emission of paired 
components versus mean strength of the secondary components
%Red to blue area ratio versus strength of the secondary component 
relative to the central component,
i.e., area ratio versus $\varepsilon$. 
Each point corresponds to one pair in one of the galaxies, as indicated in the inset -- each galaxy has a color, and the 
different symbols with same color correspond to different weak components of the same galaxy. When only one component 
exists, it tends to be the blue one (they are marked with an asterisk in the inset). 
We show unpaired components as having an area ratio equal to zero. 
Except for ML32, the ratio is consistent with one or smaller. 
There is no obvious trend in the observed scatter plot. %ratio with the strength of the component $\varepsilon$.
}
\label{blue_red}
\end{figure}

\begin{figure}
\includegraphics[width=0.5\textwidth]{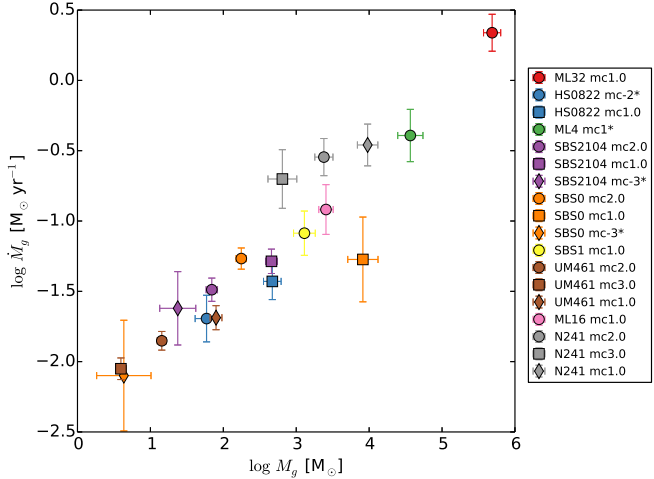}
\caption{Mass loss rate versus gas mass for the faint components observed in the wings of H$\alpha$. 
The color code and symbols have the same meaning as in the previous figure, which is indicated in the inset.
}
\label{fig:mass_loading}
\end{figure}

\begin{figure}
\includegraphics[width=0.5\textwidth]{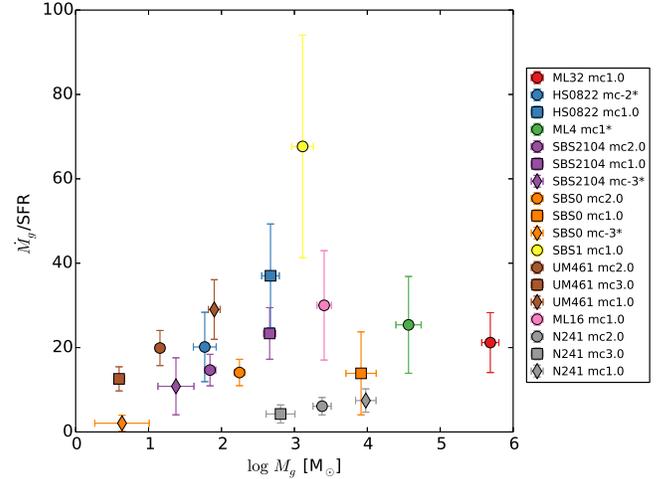}
\caption{Mass loading factor ($\dot{M}_g/{\rm SFR}$) versus gas mass, for the weak components 
observed in the  wings of H$\alpha$. All factors are larger than one, meaning that for every 
unit mass transformed in stars, several of units are swept away by the expansion.   
The color code and symbols have the same meaning as in previous figures,
which is indicated in the inset.
}
\label{fig:mass_loading2}
\end{figure}

\begin{figure}
\includegraphics[width=0.5\textwidth]{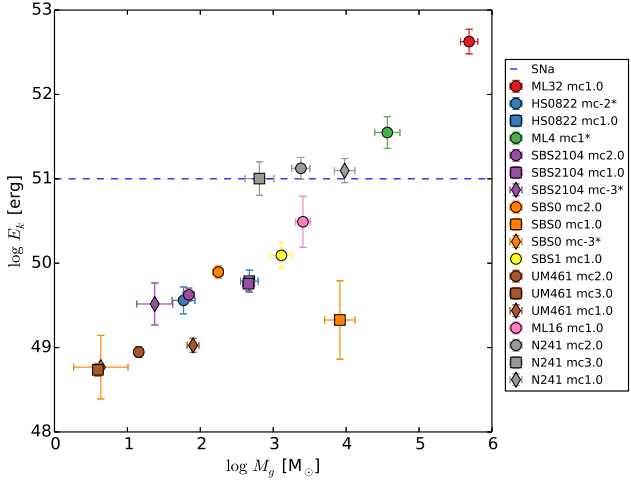}
\caption{Kinetic energy versus gas mass, for the weak components 
observed in the  wings of H$\alpha$. %They are shown in a log-log plot.
The line of $10^{51}$\,erg corresponds to the typical kinetic energy released
by a core-collapse SN, characteristic of massive stars.   
The color code and symbols have the same meaning as in previous figures,
which is indicated in the inset.
}
\label{fig:kinetic}
\end{figure}

\begin{figure}
\includegraphics[width=0.5\textwidth]{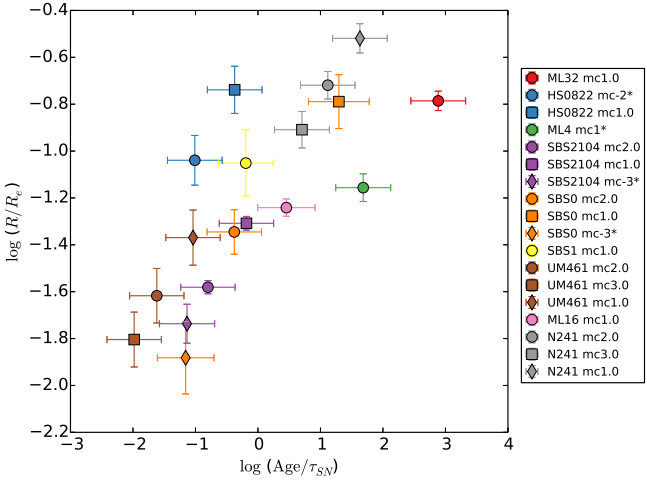}
\caption{Radius of the shell in terms of the radius of the star-forming region ($R/R_e$) versus
$Age$ of the shell relative to the expected time-lag between SN explosions ($\tau_{SN}$).
The color code and symbols have the same meaning as in previous figures,
which is indicated in the inset.
}
\label{fig:size_age}
\end{figure}

\begin{figure}
\includegraphics[width=0.5\textwidth]{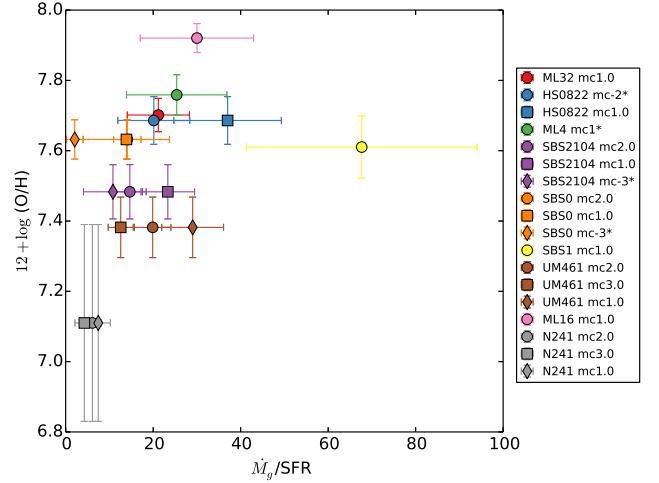}
\caption{
Mean metallicity of the star-forming region versus mass loading factor.
The color code and symbols have the same meaning as in previous figures,
as indicated in the inset.
}
\label{fig:metal1}
\end{figure}

The first significant result is the mere existence of multiple components in the wings of H$\alpha$, 
which are often paired, so that for each red component there is a blue component with a similar wavelength
shift and amplitude 
(see the example in Fig.~\ref{fig_lobes}a).
This fact discards 
spatially-unresolved, optically-thin, uniform expanding shells as the 
source of emission in the line wings, since they produce a top-hat line profile
without individual emission peaks 
\citep[e.g.,][and also Appendix~\ref{appendix}]{1987soap.conf..185Z,1994vsf..book..365C,1996ApJ...456..264T}. 
However, the two-hump emission can be caused by a non-spherically-symmetric shell, or a shell 
having internal absorption (the other alternatives are analyzed and discarded in Sect.~\ref{sec:interpretation}).
%%%
Even though the observed amplitudes are similar, the blue component of each pair tends to be larger than 
the red component. Figure~\ref{blue_red} shows the 
ratio between the area of the red and the blue components,
and in all but one case, this ratio is one or smaller, considering error bars. 
Moreover, if one of the components is missing, it tends to be the red one.
These two facts suggest a dusty expanding shell model, where the receding cap of the shell is 
obscured by the  approaching cap (see Appendix and Fig.~\ref{app:example}).  
These issues are discussed in detail in Sect.~\ref{sec:interpretation}. 
The color code employed in Figs.~\ref{blue_red}~---~\ref{fig:bh_mass}  assigns a color to each galaxy,
and then the different symbols of the same color correspond to the various pairs of
secondary components in that galaxy.

Using the equations in Sect.~\ref{mass_loss_rate}, we have estimated the mass, mass loss rate, and kinetic energy 
associated with all the weak components in the H$\alpha$ wings (Sect.~\ref{lobes}; Fig.~\ref{fig_lobes}).
The mean values, as well as the errors, are given in Table~\ref{tab:summary_lobes}.
\begin{table*}[!h] 
\caption{Parameters for the weak components in the H$\alpha$ wings\,$^a$}
\scriptsize 
\centering 
\begin{tabular}{lccccccccc}
\hline 
\noalign{\smallskip} 
ID\,$^b$    & Velocity\,$^c$  & $\varepsilon\,^d$ & red/blue\,$^e$ & $M_g$  & $R$ &  $\dot{M_g}$ & $E_k$  & Age \\ 
 & [km~s$^{-1}$] & & & [$10^{3}$\ M$_\odot$]  & [pc] &  [${\rm M}_\odot$\,yr$^{-1}$] & [$10^{51}$\ erg] & [$10^{4}$\ yr] \\ 
\noalign{\smallskip} 
\hline
\noalign{\smallskip} 
ML4-1.\,*  &  98 $\pm$ 8 & 0.031 $\pm$ 0.012 & --- & 37 $\pm$ 15 & 30 $\pm$ 4  & 0.41 $\pm$ 0.17 & 3.5 $\pm$ 1.5 & 30 $\pm$ 5 \\ 
\noalign{\smallskip} 
HS0822-1.0 &  114 $\pm$ 6 & 0.019 $\pm$ 0.005 & 0.6 $\pm$ 0.15    & 0.47 $\pm$ 0.13 & 4.9 $\pm$ 0.5  & 0.037 $\pm$ 0.011 & 0.061 $\pm$ 0.018 & 4.2 $\pm$ 0.5 \\ 
HS0822-2\,* &  -248 $\pm$ 9 & 0.0024 $\pm$ 0.0008 & ---   & 0.06 $\pm$ 0.02 & 2.5 $\pm$ 0.3  & 0.02 $\pm$ 0.008 & 0.036 $\pm$ 0.013 & 0.97 $\pm$ 0.12 \\   
\noalign{\smallskip} 
SBS0-1.0 &  16 $\pm$ 8 & 0.28 $\pm$ 0.13 & 0.12 $\pm$ 0.03   & 8 $\pm$ 4 & 8.5 $\pm$ 1.3  & 0.05 $\pm$ 0.04 & 0.02123 $\pm$ 0.02 & 50 $\pm$ 30 \\ 
SBS0-2.0 &  211 $\pm$ 0.4 & 0.0061 $\pm$ 0.0005 & 1.3 $\pm$ 0.6   & 0.18 $\pm$ 0.03 & 2.35 $\pm$ 0.13  & 0.054 $\pm$ 0.009 & 0.079 $\pm$ 0.013 & 1.09 $\pm$ 0.06 \\ 
SBS0-3\,* &  -370 $\pm$ 20 & 0.00015 $\pm$ 0.00013 & ---  & 0.004 $\pm$ 0.004 & 0.68 $\pm$ 0.2  & 0.008 $\pm$ 0.007 & 0.006 $\pm$ 0.005 & 0.18 $\pm$ 0.05 \\ 
\noalign{\smallskip} 
SBS1-1.0 & 98 $\pm$ 3&0.11 $\pm$ 0.03&0.51 $\pm$ 0.03& 1.3 $\pm$ 0.4 & 5.3 $\pm$ 0.6  & 0.08 $\pm$ 0.03 & 0.12 $\pm$ 0.04 & 5.3 $\pm$ 0.6 \\ 
\noalign{\smallskip} 
ML16-1.0 &  110 $\pm$ 40 & 0.036 $\pm$ 0.006 & 3 $\pm$ 4 & 2.6 $\pm$ 0.6 & 8 $\pm$ 0.6  & 0.12 $\pm$ 0.05 & 0.3 $\pm$ 0.2 & 7 $\pm$ 2 \\ 
\noalign{\smallskip} 
UM461-1.0 &  116 $\pm$ 3 & 0.0165 $\pm$ 0.002 & 1.2 $\pm$ 0.5 & 0.08 $\pm$ 0.015 & 1.53 $\pm$ 0.09  & 0.021 $\pm$ 0.004 & 0.011 $\pm$ 0.002 & 1.29 $\pm$ 0.09 \\ 
UM461-2.0 & 248 $\pm$ 3 & 0.00297 $\pm$ 7e-05 & 1.1 $\pm$ 0.5 & 0.014 $\pm$ 0.002 & 0.86 $\pm$ 0.04  & 0.014 $\pm$ 0.002 & 0.0088 $\pm$ 0.0013 & 0.34 $\pm$ 0.017 \\ 
UM461-3.0 & 372 $\pm$ 4 & 0.00082 $\pm$ 7e-05 & 1.2 $\pm$ 1.4  & 0.0039 $\pm$ 0.0007 & 0.56 $\pm$ 0.03  & 0.0089 $\pm$ 0.0016 & 0.0054 $\pm$ 0.0009 & 0.148 $\pm$ 0.008 \\ 
\noalign{\smallskip} 
N241-1.0 &  115 $\pm$ 4 & 0.017 $\pm$ 0.005 & 0.5 $\pm$ 0.3  & 10 $\pm$ 3 & 10.7 $\pm$ 1.2  & 0.35 $\pm$ 0.12 & 1.3 $\pm$ 0.4 & 9.2 $\pm$ 1 \\ 
N241-2.0 &  236 $\pm$ 9 & 0.0043 $\pm$ 0.0011 & 2 $\pm$ 3 & 2.4 $\pm$ 0.7 & 6.8 $\pm$ 0.6  & 0.28 $\pm$ 0.09 & 1.3 $\pm$ 0.4 & 2.8 $\pm$ 0.3 \\ 
N241-3.0 &  395 $\pm$ 7 & 0.0012 $\pm$ 0.0005 & 0.2 $\pm$ 1.4  & 0.6 $\pm$ 0.3 & 4.4 $\pm$ 0.7  & 0.2 $\pm$ 0.1 & 1 $\pm$ 0.5 & 1.08 $\pm$ 0.17 \\ 
\noalign{\smallskip} 
SBS2104-1.0  &  112 $\pm$ 2 & 0.025 $\pm$ 0.002 & 1.1 $\pm$ 0.3   & 0.46 $\pm$ 0.09 & 3.4 $\pm$ 0.2  & 0.052 $\pm$ 0.01 & 0.057 $\pm$ 0.011 & 2.96 $\pm$ 0.19 \\  
SBS2104-2.0 &  246 $\pm$ 4 & 0.0038 $\pm$ 0.0002 & 1 $\pm$ 0.5    & 0.07 $\pm$ 0.013 & 1.8 $\pm$ 0.11  & 0.032 $\pm$ 0.006 & 0.042 $\pm$ 0.008 & 0.72 $\pm$ 0.04 \\ 
SBS2104-3\,*  &  -372 $\pm$ 10 & 0.0013 $\pm$ 0.0007 & ---  & 0.024 $\pm$ 0.014 & 1.3 $\pm$ 0.2  & 0.024 $\pm$ 0.014 & 0.033 $\pm$ 0.019 & 0.33 $\pm$ 0.06 \\
\noalign{\smallskip} 
ML32-1.0 &  93 $\pm$ 9 & 0.062 $\pm$ 0.014 & 1.7 $\pm$ 0.4   & 490 $\pm$ 130 & 71 $\pm$ 6  & 2.2 $\pm$ 0.7 & 42 $\pm$ 14 & 74 $\pm$ 10 \\ 
\noalign{\smallskip} %\hline
\hline
\end{tabular} 
\begin{tabular}{l}
\noalign{\smallskip} 
$^a$~The error bars in this table represent the standard deviation across the starburst 
(velocity, $\varepsilon$, and area ratio), or errors for the value obtained\\ ~~~from
the mean H$\alpha$ line of the starburst (the rest).\\
$^b$~The number added to the galaxy name identifies each one of the pairs of components in the line profiles, with 
`1.0' representing the pair of\\ ~~~lowest velocity, `2.0' the second lowest velocity, and so on.\\ 
$^c$~Velocity of the center of the component. Negative, when there is no red counterpart to the blue component. 
In paired components, mean of\\ ~~~the absolute value of the pair.\\ 
$^d$~Area ratio between the secondary component and the main component. In paired components, 
total $\varepsilon$ of the pair.\\
$^e$~Area ratio between the red and blue emissions in the case of paired components.\\
$*$~Unpaired component.\\
\end{tabular}
\label{tab:summary_lobes}
\end{table*}
The velocities and the ratio between the secondary components and the central component, $\varepsilon$, 
are provided directly by the multi-Gaussian fit described in Sect.~\ref{lobes}. 
In the case of paired components, the velocity is the mean of the absolute value 
of the two velocities, and $\varepsilon$ is the
total $\varepsilon$ of the pair.
The estimates also depend on the temperature and electron density of the emitting plasma,
as well as on the relative thickness of the shell, $\Delta R/R$, and the filling factor, $f$.

The electron temperature is taken from the data used to determine oxygen abundance 
in Paper~I. Specifically, we use the mean over the main star-forming region of the galaxy.
The electron density has been inferred from the ratio of the emission lines [SII]$\lambda$6716 and [SII]$\lambda$6731,
following the standard procedure \citep[e.g.,][]{1974agn..book.....O},
using the mean value over the whole star-forming region.
These two lines are present 
within the spectral range observed with the WHT.  Electron densities and temperatures are listed in Table~\ref{tab:mean_data}.
As for the thickness of the shell, we use $\Delta R/R=0.3$ (see Sect. 3.3). The filling factor, $f$, is assumed 
to be one.

%%%
The velocities of the weak secondary components are typically in excess of 100~km~s$^{-1}$ (see Table~\ref{tab:summary_lobes}).
These velocities are very large compared with the rotational and turbulent velocities 
of the galaxies (Sect.~\ref{sect:rot_turb}), and, hence, are almost certainly larger than the escape velocity 
of the  galaxy, expected to be of the order of the rotational 
plus turbulent velocity in the disk \citep[e.g.,][]{2008gady.book.....B}.
Assuming that the components in the wings correspond to outflows in expanding shells, it is still unclear 
whether or not the involved mass will finally escape from the galaxy. It all depends on the ISM 
of the host galaxy, and the pressure it exerts on the 
expanding shell. The issue is analyzed in Sect.~\ref{sec:interpretation}, with the conclusion that the gas will likely escape.
The mass involved in these motions is given in Table~\ref{tab:summary_lobes} and summarized in Fig.~\ref{fig:mass_loading}.
The masses are in the range between $10\,{\rm M}_\odot$ and $10^5\,{\rm M}_\odot$, 
and they represent a few percent of the total gas mass in the HII region 
(the parameter $\varepsilon$ in Table~\ref{tab:summary_lobes} gives the 
flux ratio between the secondary and the main emission lobe,
and it has been taken as a proxy  for the mass ratio; see Eqs.~[\ref{epsilon}] and [\ref{massloss2}]).
These masses are very small compared to the stellar masses and the dynamical masses
of the entire galaxies (Tables~\ref{table_obs} and \ref{table_rotation}). 

Figure~\ref{fig:mass_loading} shows that the mass loss rate scales with the 
gas mass, since the velocities and radii of the different components and galaxies 
are similar. The weak components have associated mass loss rates 
between $10^{-2}$ and 1~M$_\odot$~yr$^{-1}$. They are very significant compared
with the SFRs, as is attested in Fig.~\ref{fig:mass_loading2}, which shows the 
mass loading factor, $\dot{M}_g/{\rm SFR}$. Most mass loading factors are in excess of one, 
meaning that for every mass unit transformed into stars, several mass units are swept away. 
Whether this mass escapes the HII region, and later on, the galaxy, we cannot know from 
our observations.  However, cosmological numerical
simulations predict large mass loading factors in low mass galaxies 
\citep[10 is not uncommon; see, e.g.][]{2011MNRAS.417.2962P,2012MNRAS.421...98D,2012MNRAS.420..829O,
2014ApJ...780...57B,2016MNRAS.455..334T}.
%Therefore, the mass involved in the observed motions conforms to the theoretical predictions.
Mass loading factors observed in high redshift objects are generally lower 
than the ones we infer \citep[say, of the order of 2, e.g., ][]{2012ApJ...761...43N,2012ApJ...760..127M},
likely because they refer to massive galaxies ($\log(M_\star/M_\odot) > 10$). However,
it is not uncommon to find factors up to 10 in local dwarfs \citep[e.g.,][]{1999ApJ...513..156M,2005ARA&A..43..769V}. 
A caveat is in order, though. We are assuming the density of the shell to be the 
same as the mean density of the star-forming region. If the expanding shell builds up mass  
by sweeping gas around the explosion site, its density is expected to be higher than the mean 
density of the region. This increase of density would reduce our mass loading factor estimates
(see Eq.~[\ref{massloss2}]).

Figure~\ref{fig:kinetic} shows the kinetic energy associated with the weak
components. As with the mass loss rate, it scales with the mass 
of gas that is involved in the motion. Figure~\ref{fig:kinetic} also shows the 
energy typically  associated with a core-collapse SN explosion, of the order of $10^{51}$\,ergs  
\citep[e.g.,][]{1999ApJS..123....3L}. The energies involved in the observed  motions are usually around this 
value, albeit smaller; ML32 is the only exception. Therefore, energy-wise, the components may be tracing shells produced 
by individual SN explosions.  ML32 would require some 10 SN exploding simultaneously. 

Depending on the IMF (initial mass function), population synthesis models 
predict a SN rate between 0.02 and 0.005~yr$^{-1}$ for a constant 
SFR=1~M$_\odot$~yr$^{-1}$ \citep{1999ApJS..123....3L}. Using a value in between
these two extremes, say 0.01~yr$^{-1}$/$({\rm M}_\odot {\rm yr}^{-1})$, the SFRs of our star-forming regions 
(Table~\ref{tab:mean_data}) allow us to estimate the expected SN rates assuming the observed SFRs 
to be constant. 
The predicted SN rate turns out to be between $10^{-3}$ 
and  $10^{-5}~{\rm yr^{-1}}$, which corresponds to a time-lag between SN explosions 
($\tau_{SN}$, defined as the inverse of the SN rate) from 1000 yr to 0.1 Myr.
These rates suffice to maintain the observed signals.
%, since the $Age$ associated 
%with the shells (Table~\ref{tab:summary_lobes}) is often
%longer than the time-lag. Figure~\ref{fig:size_age} provides 
%the ratio between the $Age$ of the shell  and  the time-lag between explosions
Figure~\ref{fig:size_age} provides 
$Age/\tau_{SN}$ for individual shells. $Age > \tau_{SN}$ in half of the cases, 
which implies expecting more than one SN signal at a time, as we observe. 
%, and 
%\modified{and both time-scales are similar although with very large scatter in the ratio.}

Ages are estimated from the measured velocity and the estimated radius of the shell (Eq.~[\ref{eq:age_shell}]).
These radii are small compared to the size of the star-forming regions. Figure~\ref{fig:size_age} 
shows the radius relative to the effective radius of the region, with the typical ratio being around 10\,\% or smaller.
Therefore, it is important to keep in mind, for interpreting the weak emission components in the 
wings of H$\alpha$, that there are one or several (probably non-concentric) expanding shells 
embedded in a large star-forming region. This substructure is common in giant HII regions, such as 30 Doradus 
in the large Magellanic cloud \citep[e.g.,][]{1984MNRAS.211..521M,2013AJ....146...53S,2015MNRAS.447.3840C}.
Shells and holes appear in Hubble Space Telescope (HST) images of the large starburst in Kiso~5639, a tadpole galaxy 
very similar to the XMPs studied here \citep{2016arXiv160502822E}.

%%%%%%%%
We do not see a dependence of the properties of the weak emission-line components 
on the mean metallicity  of the starburst, except perhaps for the mass loading factor.
Figure~\ref{fig:metal1} shows the scatter plot of metallicity versus mass loading factor,
and it hints at an increase of the factor with increasing metallicity. 
The trend may be coincidental, due to the small number of points 
in the plot. 
However, it may also indicate the contribution of stellar winds
to the energy and momentum driving the observed expansion. 
The winds of massive stars can provide a total kynetic energy comparable 
to the energy released during their SN explosions \citep[e.g.,][]{2016MNRAS.456..710F}, 
and these radiation-pressure driven winds have mass loss rates that 
increase with increasing metallicity \citep[e.g.,][]{2015sf2a.conf..343M}.
The metallicities represented in Fig.~\ref{fig:metal1} correspond to the  
metallicity averaged over the starburst and, thus, due to the variation of metallicity
across the star-forming region, their value is slightly higher than the values quoted in Paper~I.
ML16, the point of highest metallicity in Fig.~\ref{fig:metal1}, is also the galaxy with no
detected metallicity inhomogeneity in Paper~I.

%%%%%%%%%%%%%%%%%%%%%%%%%%%

\subsection{Other interpretations}\label{sec:interpretation}

One can think of several alternatives to explain the existence of multiple paired emission-line components in 
the wings of H$\alpha$. 
The previous section analyzes the possibility of being produced by
spatially-unresolved expanding shells 
driven by SN explosions. 
Here we discuss other options, such as 
(1) spatially-unresolved rotating structures, 
(2) spatially-resolved expanding shells, 
(3) bipolar outflows,  
(4) blown out spherical expansion leading to expanding rings, 
and (5) spatially-unresolved expanding shells driven by BH accretion feedback.
%the alternative possibilities, including (1) for completeness. 

%(1) {\em Spatially-unresolved expanding shells driven by SN explosions.} 
%This possibility is  analyzed in Sect.~\ref{sec:mult1}, where we find that individual 
%SNa explosions produce enough energy to drive the observed motions. The observed SFR leads to 
%a SN rate that is consistent with the age of the shells. One potential problem for this option is 
%that  the spatially integrated emission of an optically thin shell does not present individual
%peaks, but a top-hat line profile. %  \citep[e.g.,][]{1987soap.conf..185Z,1994vsf..book..365C,1996ApJ...456..264T}.  
%However, if the shell has some internal extinction, then the line 
%center photons are preferentially absorbed, which produces an emission line with
%two peaks. We show in Appendix~\ref{appendix} that two-horned profiles emerge even  
%for the weak dust extinction expected in XMPs. %\citep[][]{2016ApJ...819..110S}. 
%Double horn profiles also result in rapidly expanding optically thick shells 
%\citep[][]{1994vsf..book..365C}, but it is unclear
%whether this case is important for our context because the H$\alpha$
%line opacity is not expected to be large:  most H is ionized, and even 
%the neutral H is expected to be in the ground 
%state\footnote{\red{This must be proven, eventually.}}.
%

\begin{figure}
\includegraphics[width=0.5\textwidth]{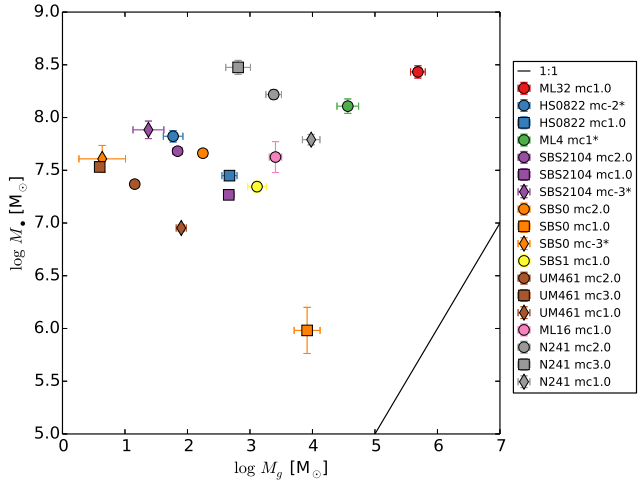}%jpg}\\
\caption{Mass of the central object versus gas mass if the pairs of faint emission components are 
interpreted as caused by a gaseous disk orbiting around a central massive object. The solid line represents 
the one-to-one relationship.  The masses of the required central objects turn out to be  unrealistically large, 
which disfavors this interpretation.
The color code and symbols have the same meaning as in previous figures,
which is indicated in the inset.
%The inset gives the color code for the different symbols, which have the same meaning as in previous figures.
}
\label{fig:bh_mass}
\end{figure}

(1) {\em Spatially-unresolved rotating disk-like structures.}
Rotation naturally produces double-horned  profiles \citep[e.g.,][]{1964AJ.....69..521E}. Therefore,
one of the possibilities to explain the existence of blue and red paired emission peaks 
is the presence of a gaseous structure rotating  around a massive object.  
%The approaching and receding parts of the disk-like structure would be responsible for the blue and 
%the red peaks, respectively. %\citep[e.g.,][]{1964AJ.....69..521E}.  
The mass of the required central object can be estimated from 
the size of the rotation structure and the velocity (Sect.~\ref{sec:mass_bh}). 
Figure~\ref{fig:bh_mass} shows that the
%the mass of the putative central object versus the gas mass. The 
mass of the central object needs to be around three orders of magnitude larger 
than the gas mass involved in the motions, with central masses, $M_\bullet$, ranging from 
$10^6$--$10^8$\,M$_\odot$. 
%The radius of the gaseous disk, on the other 
%hand, would have to be similar to the radius estimated for the shell (Table~\ref{tab:summary_lobes}), 
%and, hence, in the range between 1 and 100~pc. The mass of the required central object 
This mass
turns out to be of the order of the stellar mass of the whole galaxy (Table~\ref{table_obs}), 
which rules out the stellar nature of the required central concentration. 
If, on the other hand, it were a BH, then it would have to have $M_\bullet\sim M_\star$, which is completely out 
of the so-called Magorrian-relationship between the central BH mass and the 
stellar mass  %(of the spherical component) 
of the galaxy, for which
$M_\bullet/M_\star\sim 10^{-3}$ \citep[e.g.,][]{2013ARA&A..51..511K}. 
Thus, the exotic nature of the required central mass disfavors the 
interpretation  of the weak emission in the wings of H$\alpha$ as caused by a 
disk orbiting around a massive central object, such as a BH.

(2) {\em Spatially-resolved expanding shells.}
The inner part of a spatially resolved expanding shell also produces 
double peak emission-line profiles, since the outer parts of the shell, contributing
to the line core, are excluded \citep[e.g.,][]{1987soap.conf..185Z}.  
%However, w
We often see several pairs of emission peaks (Fig.~\ref{fig_lobes}a), therefore, 
this interpretation requires the existence of several resolved concentric shells, 
which is unrealistic since the first shell would have already swept the ISM around
the center of the expansion.
%which is an unrealistically ordered scenario.
Moreover, the size of the shell is estimated  to be
%from the observations in a way
%that is fairly independent of details of the geometry; given the mass of gas 
%inferred from the H$\alpha$ emission flux, and the density (inferred from 
%emission line ratios), the size of the emitting structure is set. This 
%size turns out to be much smaller than the star formation region (Fig.~\ref{fig:size_age})
%and, thus, 
smaller than the typical spatial resolution, so that the expanding regions cannot be spatially resolved.

(3) {\em Bipolar outflows.}
Insofar as they comply with the energetics required to explain the observed motions,
bipolar flows cannot be easily discarded. They create two-horned profiles,
and several of them coexisting in the resolution element could explain 
the presence of multiple paired components. In fact, bipolar flows are extreme cases
of non-spherically-symmetric shells, where double-horned line profiles
appear naturally \citep[e.g.,][]{1999MNRAS.307..677G}. 

(4) {\em Blown out spherical expansion leading to expanding rings.}
If the radius of the shell exceeds the scale height of the disk of the 
galaxy, it may break up, leading to a ring-like expanding 
structure that produces two-horned profiles when observed edge-on.
%structure enclosed within the galaxy disk. Expanding rings 
% naturally produces two-horned line profiles. %However, 
This possibility can be discarded since the ring would have to be as large as
the disk thickness, which is incompatible with the small size of the expanding 
structures (Table~\ref{tab:summary_lobes}).
%by resorting to the argument on the small size of the expanding structure.
%, already posed above. For a
%shell to be blown out, its size would have to be larger than the galaxy disk height, 
%which is not consistent with the inferred sizes of the expanding structures, which are only a few 
%pc (Table~\ref{tab:summary_lobes}).

(5) {\em Spatially-unresolved expanding shells driven by BH accretion feedback.}
Energywise, intermediate mass BHs of $10^5~{\rm M}_\odot$
can also power the expanding shells that we characterize in Sect.~\ref{sec:mult1} 
(see Eq.~[\ref{eq:bh_mass2}]). However, the observations also disfavor this option for a number of reasons. 
It is unclear how a primordial BH seed can coincide spatially with a star-forming region unless 
there are many such BHs lurking in the galaxy, or the BH is born in situ.
%
%The BH masses of interest are around . They 
$10^5~{\rm M}_\odot$ BHs
can be produced by direct collapse in the early universe, but this explanation can be discarded, since they
are very rare \citep[e.g.,][]{2010MNRAS.408.1139V} and many of them are needed.
% detection of these BHs in dwarf galaxies  has been largely elusive \citep[e.g.,][]{2010MNRAS.408.1139V}. 
On the other hand, $100~{\rm M}_\odot$ BHs are expected to be more common, left as
remnants from population III stars formed in the early universe or even today 
\citep[e.g.,][]{2015ApJ...801L..28K}. However, this alternative
is also very unlikely, since HII regions 
are short-lived \citep[a few tens of Myr; e.g.,][]{1999ApJS..123....3L}, 
and a BH requires several hundred Myr of continuous feeding to grow from $100$ to $10^5~{\rm M}_\odot$ 
\citep[see, e.g.,][]{2010A&ARv..18..279V}.
%
%In addition 
%to the difficulties posed by the mere existence of BHs in the star-forming 
%regions, the fact that sometimes we observe several pairs of emission peaks would
%imply the existence of several concentric shells, a scenario difficult to
%reconcile with the BH scenario, since the first shell would have already swept the ISM around the BH. 
%Alternatively,
%if the shell mass is ejected from the BH accretion disk, then it would be difficult to explain 
%ejecta as massive as $10^5~{\rm M}_\odot$. 
Finally, accreting BHs are expected to be luminous X-ray sources. Five of our XMPs
were observed with the X-ray satellite Chandra: HS0822, SBS1, N241, SBS2109,
and UM 461 \citep[][]{2011ApJ...741...10K,2013ApJ...769...92P}.
Only one of them shows a signal above the noise, namely, SBS1, with an X-ray luminosity around 
$10^{39}~{\rm erg\,s}^{-1}$. This level of emission is consistent with %X-ray emission from 
young SNe \citep[e.g.,][]{2004A&A...418..429F,2006A&A...460...45G,2012MNRAS.419.1515D}.
%
%In summary, the feedback from BHs of $10^5~{\rm M}_\odot$ is unlikely to explain our observation.

%%%%%%%%%%%%%%%%%%
\section{The fate of the swept out material}\label{sec:mydiscussion}

Independently of the details on whether the shells are symmetric, or if they have internal extinction, 
the expanding shell 
model appears to be the best explanation to reproduce the weak emission features observed in 
the H$\alpha$ wings. One of the consequences 
of this interpretation is that the mass loading factor that we infer may be reflecting the feedback of 
the star formation process on the medium in, and around, the galaxy. 
The measured mass loading factor is very large,  of the order of ten or larger 
(Fig.~\ref{fig:mass_loading2}), which implies that for every solar mass of gas transformed 
into stars, at least ten times more gas is swept away from the region. How far this 
swept gas goes is impossible to tell from the data alone. However,  a significant fraction may 
escape from the galaxy to the CGM and even the intergalactic medium (IGM). 
The speeds, larger than 100~km\,s$^{-1}$
and up to 400~km\,s$^{-1}$ (Table~\ref{tab:summary_lobes}), likely  exceed the escape velocity of the galaxy. 
The final fate of the gas depends
on whether the gas pressure in the ISM of 
the galaxy suffices to slow down and confine the expansion. Even though we do not know the gas pressure of the 
ISM, it must be similar to the turbulent pressure in the star-forming region, so that
the region is neither expanding nor shrinking considerably (the moderate global expansion observed in some 
cases is analyzed in the paragraph at the end of the section). 
The turbulent pressure scales with the square of the 
turbulent speed, and so does the pressure exerted by the shell on the ISM \citep[e.g.,][]{1951RSPSA.210...18C,1987A&A...172..293B}. 
The measured turbulent velocities are of the order of 25~km\,s$^{-1}$
(see Table~\ref{tab:mean_data}, and keep in mind that the quoted values are FWHM velocities rather than velocity dispersions),
hence, provided that the HII region and the shell have similar densities, the 
ISM will not be able to confine a expanding shell with a speed of a few hundred km\,s$^{-1}$. 
A significant part of the matter in the shell may eventually escape from the galaxy. 
%In other words,
%the large mass loading factors that we measure may well represent the mass loading factors
%of the entire galaxy.
\citet{1999ApJ...513..156M} made a qualitative estimate of the conditions for starburst-driven outflows
to escape, finding that the gas flows out from  galaxies with rotation smaller than 130~km\,s$^{-1}$,
i.e., from our XMPs. 

Under the hypothesis of expanding shell, it is possible to compute the 
column density to be expected if the shell material reaches the CGM.  
Assuming mass conservation, and if the shell geometry is maintained, the 
typical column density when the shell expands to a radius $R'$ is, 
\begin{equation}
N_H\simeq 7.9\times 10^{14}\,{\rm cm^{-2}}\,{{R}\over{10\,{\rm pc}}}\,
{{n_e}\over{10^2\,{\rm cm^{-3}}}}\,\Big[{{R/R'}\over{10^{-3}}}\Big]^2,
\label{eq:column_density}
\end{equation}
where the electron density, $n_e$, and the radius, $R$, are those measured in the star-forming regions. 
Note that for the typical sizes and densities of our shells, and with expansion factors  of the order of 
$10^3$ that transform pc-size shells into intergalactic size structures (from $R\simeq 10$\,pc to $R'\simeq 10$\,kpc), 
the column densities predicted by Eq.~(\ref{eq:column_density})
are of the order of $10^{15}\,{\rm cm^{-2}}$. This gas  would be undetectable with current instrumentation, even if it 
were fully neutral. For example, a high-sensitivity survey such as 
HIPASS sets the limit at  $4\times 10^{17}\,{\rm cm}^{-2}$ \citep{2004MNRAS.350.1195M}, 
and values of $10^{17}\,{\rm cm}^{-2}$ will be achieved by SKA after a 1000~hour integration 
\citep{2015arXiv150101211D}.
Moreover,  at these low column densities the intergalactic ultraviolet (UV) background
will keep the gas fully ionized, which makes it even more elusive. 
How long would it take for the gas to reach the CGM? Assuming that the original
velocities are maintained, the time to reach the CGM is the expansion factor, $R'/R$,
times the $Age$ in Eq.~(\ref{eq:age_shell}). Given the $Age$ in Table~\ref{tab:summary_lobes}
and $R'/R\simeq 10^3$,  the time-scale is of the order of 10~Myr, which is short compared to the 
time-scale for galaxy evolution. In summary, the shells that we observe in the star-forming regions are expected to 
reach the CGM, and may do it in a reasonable time-scale. 
Unfortunately, they would be undetectable with current instrumentation unless 
the expansion is more moderate than expected according to Eq.~(\ref{eq:column_density}). 
%
%For example, Eq.~(\ref{eq:column_density}) assumes that the individual  shells will not interact with each other during expansion. 
%However, in reality, they will, hampering the expansion of parts of the shells, and thus reducing the mass loss rate with 
%respect to those in Fig.~\ref{fig:mass_loading}. This effect does not necessarily 
%reduce the gas column density, since the interaction may collimate the outflows 
%\citep[e.g.,][]{2003ApJ...597..279T}, leading to a more moderate expansion than the extreme 
%3D dilution assumed in Eq.~(\ref{eq:column_density}). 

We show in Sect.~\ref{sect:rot_turb} that some of the star-forming regions may be undergoing a 
global expansion, with a moderate velocity of around 15~km\,s$^{-1}$. The mass loss rate associated with this
global expansion, $\dot{M}_G$, can also be estimated using Eq.~(\ref{massloss}), except that this time
the global mass of the starburst, $M_G$, and the radius of the starburst, $R_e$, have to be 
considered.  A simple substitution renders $\dot{M}_G$ in terms of the mass loss rate inferred
for one of the weak components in the wings of H$\alpha$, i.e.,
\begin{equation}
%{{\dot{M}_G}\over{\dot{M}_g}}=\simeq \varepsilon^{-1}\,{{R}\over{R_e}}\,{{15~{\rm km\,s}^{-1}}\over{v_{\rm out}}},
\dot{M}_G/\dot{M}_g\simeq \varepsilon^{-1}\,(R/R_e)\,(15~{\rm km\,s}^{-1}/v_{\rm out}),
\end{equation}
where we have assumed $\Delta R/(R_e/3)\simeq R/R_e$, and $M_g/M_G\sim \varepsilon$. Using 
typical values for these parameters, $\varepsilon\simeq 5\times 10^{-3}$ (Table~\ref{tab:summary_lobes}), 
$v_{\rm out}\simeq 150~{\rm km\,s}^{-1}$ (Table~\ref{tab:summary_lobes}),
and $R/R_e\sim 5\times 10^{-2}$ (Figure~\ref{fig:size_age}), one 
finds that the two mass loss rates are similar, i.e.,  
\begin{equation}
\dot{M}_G/\dot{M}_g\sim 1.
\end{equation}
One can repeat the exercise for the kinetic energy involved in the global expansion (Eq.~[\ref{eq:kinetic}]), which  
scales with the mass ratio and the square of the expansion velocities. Using 
the typical values given above, the two kinetic energies turn out to be similar too.
This rough estimate leads to the conclusion that the mass loss rate and the 
kinetic energy involved in the global expansion  of the star-forming region are similar to the values 
inferred for the emission features observed in the wings of H$\alpha$.  
The question arises as to whether the global expansion unbounds the gas in the starburst, and so may
contribute to the global mass loss of the XMPs.  As was the case with the small expanding shells discussed above, 
the answer to the question is very uncertain, and cannot be  addressed resorting to our observational data alone. 
The fate of the gas depends on the unknown value of the ISM pressure. In contrast with the small shells, 
the expansion velocities are significantly smaller than the expected escape velocity. Moreover, the turbulent 
velocity of the galaxies is typically 40--50 km\,s$^{-1}$, and, hence, comparable, but larger, than the turbulent 
velocity in the star-forming region (see Fig.~\ref{fig:dispersion}). 
%Only if the density in the ISM of the XMPs 
%is much lower than the density in the starburst, may the starburst be overpressurized; 
%the moderate observed expansion can then drive the gas all the way to the CGM.
Only if the density in the ISM of the XMPs is much lower than the density in the starburst, the 
starburst may be overpressureized and the mild observed expansion can drive gas through the ISM to large 
distances.
%Expansion perpendicular to the disk will be easier because the density is lower there, and then 
%the outflow is likely to be bipolar. 
%%%%%%%%%%%%%%%%%%%%%%%

\section{Conclusions}\label{conclusions}

The XMP galaxies analyzed in Paper~I are characterized  by having a large star-forming 
region with a metal content lower than the rest of the galaxy. The presence of such chemical 
inhomogeneities suggests cosmological gas accretion as predicted in numerical simulations
of galaxy formation. Cosmological gas accretion induces off-center giant star-forming clumps that 
gradually migrate toward the center of the galaxy disks \citep{2010MNRAS.404.2151C,2014MNRAS.443.3675M}.
The star-forming clumps may be born in-situ or ex-situ. In the first case, the accreted 
gas builds up the gas reservoir in the disk to a point where disk instabilities set in and trigger star 
formation. In the second case, already formed clumps are incorporated into the disk. They may come with stars 
and dark matter, and thus, they are often indistinguishable from gas-rich minor mergers  
\citep{2014MNRAS.443.3675M}.
Since they feed from metal-poor gas, the star-forming clumps 
are more metal-poor than their immediate surroundings \citep[][]{2016MNRAS.457.2605C}.  
In order to complement the chemical 
analysis carried out in Paper~I, this second paper studies the kinematic properties of the 
ionized gas forming stars in the XMP galaxies. We study
nine objects (Table~\ref{table_obs}) with a spectral resolution of the order of 7500, equivalent to
40~km\,s$^{-1}$ in H$\alpha$.

Most XMPs have a measurable rotation velocity, with a rotation amplitude which is only a few tens of km\,s$^{-1}$ 
(Fig.~\ref{fig:rot_curve}). All rotation curves present small-scale velocity irregularities, often with an amplitude
comparable to the velocity gradient across the whole galaxy. 
On top of the large-scale motion, the galaxies also present turbulent motions 
of typically 50~km\,s$^{-1}$ FWHM (see Fig.~\ref{fig:dispersion} and Table~\ref{tab:mean_data}), and, therefore, larger
than the rotational velocities. 

Observations suggest that the main star-forming region of the galaxy (Fig.~\ref{fig:all_images})
moves coherently
%is a kinematically distinct entity 
within the host galaxy. The  velocity is constant (Fig.~\ref{fig:disp}), so that the region moves as a single 
unit within the global rotation pattern of the galaxy. In addition, the velocity dispersion 
in some of these large star-forming clumps increases toward the center-side of the galaxy. 
The excess velocity dispersion on the center-side may indicate an intensification of the turbulence of the 
gas that collides with the ISM of the host galaxy, as if the clumps were 
inspiraling toward the galaxy center.
This migration to the galaxy center is expected from tidal forces  acting upon massive gas clumps
\citep[see, e.g.,][]{2008ApJ...688...67E,2012ApJ...747..105E,2008A&A...486..741B,2016MNRAS.457.2605C}, 
and the large starbursts in our XMPs 
may be going through the process at this moment. 
In some other cases, the velocity dispersion has a local maximum at the core 
of the star-forming region, which suggests a global expansion of the whole
region, with a moderate speed of around 15~km\,s$^{-1}$. Sometimes the 
velocity dispersion presents a minimum at the core, which may reflect a past global 
expansion that washed out part of the turbulent motions.

We find no obvious relationship between the kinematic  properties and the metallicity drops, 
except for the already known correlation between the presence of a starburst and the decrease in metallicity. 
We do find, however, a significant lack of correlation between the metallicity and the N/O ratio
(Fig.~\ref{fig:oh_no}), which is consistent with the gas accretion scenario. If the accretion
of metal-poor gas is fueling the observed starbursts, then the fresh gas reduces 
the metallicity (i.e., O/H), but  it cannot modify the pre-existing ratio between metals
\citep[e.g.,][]{2010ApJ...715L.128A,2012ApJ...749..185A},  forcing the N/O ratio to be similar 
inside and outside the starburst.

The H$\alpha$ line profile often shows a number of faint 
emission features, Doppler-shifted with respect to the central component.
Their amplitudes are typically a few percent of the main component (Fig.~\ref{fig_lobes}),
with the velocity shifts being between 100 and 400~km\,s$^{-1}$ (Table~\ref{tab:summary_lobes}).
The components are often paired, so that red and blue peaks, with similar amplitudes
and shifts, appear simultaneously. The red components tend to be slightly fainter, however. 
Assuming that the emission is produced by recombination of H, we have estimated the 
gas mass in motion, which turns out to be in the range between 10 and $10^5~{\rm M}_\odot$ 
(Table~\ref{tab:summary_lobes} and Fig.~\ref{fig:mass_loading}).  Assuming that the Doppler shifts are produced 
by an expanding shell-like structure, we infer a mass loss rate between $10^{-2}$ and a few ${\rm M}_\odot\,{\rm yr}^{-1}$.
Given the observed SFR, the mass loss rate yields a mass loading factor (defined as 
the mass loss rate divided by the SFR) typically in excess of 10. Large mass loading 
factors are indeed expected from numerical simulations 
\citep[e.g.,][]{2011MNRAS.417.2962P,2012MNRAS.421...98D,2012MNRAS.420..829O},
and they reflect the large feedback of the star formation process on the surrounding medium. 
Numerical simulations predict that low mass galaxies are extremely inefficient in using their gas, 
most of which returns  to the IGM without being processed through the stellar mechanism. Since the 
measured expansion velocities exceed by far the rotational and turbulent velocities, 
we argue in Sect.~\ref{sec:interpretation} that the gas involved in the expansion is bound to escape from 
the galaxy disk, reaching the CGM and possibly the IGM. Therefore, the large mass loading factors that we 
measure support and constrain the predictions from numerical simulations. 
We have estimated the H column density to be expected when the shell material reaches the CGM, and 
it turns out to be around $10^{15}$~cm$^{-2}$. Even if all the H were neutral, this is undetectably small 
with present technology. Moreover, at these low column densities, all H is expected to be 
ionized by the cosmic UV background. Even if the IGM of the XMPs were filled by expanding 
shells from previous starbursts, they would be extremely elusive observationally.  

From an energetic point of view, the observed motions involve energies between $10^{49}$ and 
$10^{52}$~erg, which are in the range of the kinetic energy released by a core-collapse 
SN ($\sim 10^{51}$~erg). The low-end energy requires that only part of the SN kinetic
energy drives the observed expansion, whereas the high-end energy requires
several SNe going off quasi-simultaneously. The observed SFRs allow for the 
required number of SNe, and the estimated age of the shells ($10^4$--$10^6$~yr; Table~\ref{tab:summary_lobes})
is also consistent with the expected SN rate. 
Other alternatives to explain the weak emission peaks in the wings of H$\alpha$, 
like BH driving the motions or the expansion,  can be discarded using physical arguments 
(Sect.~\ref{sec:interpretation}). 

Some of the data suggest a moderate global expansion of the star-forming regions (Sect.~\ref{sec:mydiscussion}). 
Even if the expansion velocity is small (around 15~km\,s$^{-1}$), 
the motion involves the mass of the whole region, and, hence, the associated mass loss rate 
and kinetic energy turn out to be similar to those carried away by the fast-moving secondary 
components. Thus, global expansion may also be important in setting up the mass and energy 
budgets associated with star formation feedback.

High redshift clumpy galaxies are probably growing through a cosmological gas accretion 
process similar to the one experienced by XMPs. However, the physical conditions at 
high and low redshifts differ (e.g., the gas accretion rates are larger at high redshifts), so do 
the observational biases affecting XMPs and high-redshift clumpies (e.g., low-mass galaxies are
undetectable at  high redshift). Therefore, addressing the issue of whether XMPs are low-mass 
analogues of high-redshift clumpies is complicated, and so goes beyond the scope of our work. 
However, because of their probable common physics, it is interesting to point out some of 
the similarities and differences of the two samples.
As in the case of XMPs, some clumpy galaxies show metallicity drops associated with 
enhanced star-formation activity \citep[][]{2010Natur.467..811C,2014A&A...563A..58T}.
High redshift galaxies are significantly more massive than XMPs. Using as reference 
for high redshift galaxies (1.3--2.6) the SINS\footnote{SINFONI Integral Field Spectroscopy of redshift two 
Star-forming Galaxies.} sample \citep{2009ApJ...706.1364F}, 
the stellar mass of the high-redshift objects ($10^{9.5}$--$10^{11}$\,$M_\odot$)
is three orders of magnitude larger than the mass of XMPs 
($10^6$--$10^{8.5}$\,$M_\odot$; Table~\ref{table_obs}).
This difference in stellar mass shows in dynamical mass too,
with rotational velocities between 100 and 260\,km\,s$^{-1}$ at high redshift. 
We cannot directly compare the rotational velocities of XMPs with these values 
since we cannot corrected for inclination, however, the uncorrected values 
go from zero to 30\,km\,s$^{-1}$ (see Fig.~\ref{fig:rot_curve}).
The velocity dispersion in XMPs ranges from 10 and 20\,km\,s$^{-1}$
(corresponding to FWHM between 25 and 50\,km\,s$^{-1}$; see Fig.~\ref{fig:dispersion}),
whereas it is between 50 and 200\,km\,s$^{-1}$ at high-redshift 
\citep{2009ApJ...706.1364F}.
The high redshift objects have a ratio between  rotational velocity and velocity 
dispersion between 0.5 and one \citep{2009ApJ...706.1364F}. XMPs seem to have 
similar ratios, even though we cannot be more precise since the measured
rotation in XMPs is affected by inclination. 
The total SFRs are very different 
(10 -- 500\,$M_\odot$\,yr$^{-1}$ at high redshift and 
0.001 -- 0.1\,$M_\odot$\,yr$^{-1}$ in XMPs), 
but the surface star-formation rates are not;
they go from 0.01 to 10\,$M_\odot$\,yr$^{-1}$\,kpc$^{-2}$ at high redshift 
\citep[e.g.,][]{2016MNRAS.tmp.1418F} 
and from 0.005 to 0.8\,$M_\odot$\,yr$^{-1}$\,kpc$^{-2}$ in XMPs (Paper~I). 
Mass loading factors are one order of magnitude larger in XMPs than those 
inferred for high-redshift clumpy galaxies 
\citep[Sect.~\ref{multiple_comp}; see also ][]{2012ApJ...761...43N}, a difference 
that we attribute to their very different masses. However, the differences 
of size between the expanding regions of clumpies and XMPs may also 
play a role (see Eq.~[\ref{massloss}]).
Local UV bright clumply galaxies \citep{2012ApJ...750...95E} are often XMPs with metallicity drops 
\citep{2013ApJ...767...74S}, and we know that their star-forming clumps resemble those in 
high-redshift cumplies in terms of total SFR, clump mass, and stellar mass surface 
density \citep{2013ApJ...774...86E}.

%\begin{itemize}
%\item[-] What about the lack of rotation in some targets?
%Is it the signal expected from face on disks?
%\end{itemize}

%%%%%%%%%%%%%%%%%%%%%%%%%%%%%%%%%%%%%%%%%%%%%

%%%%%%%%%%%
\acknowledgements

Thanks are due to R. Amor\'\i n, Y. Ascasibar, A.~B.~Morales-Luis, 
P. Papaderos,  and J. V\'\i lchez, that participated in the target 
selection and in some of the observing time proposals that rendered
the spectra analyzed in the paper.
We also thank an anonymous referee whose comments help improving the 
readability of the paper.
%
% for enlightening discussions and suggestions on various aspects of the work. 
%
This article is based on observations made with the WHT operated by the ING in the 
Observatorio del Roque de los Muchachos, La Palma, Spain.
This work has been partly funded by the Spanish Ministry of Economy and 
Competitiveness, projects {\em Estallidos} AYA2013--47742--C04--02--P and
AYA2013--47742--C04--01--P. %\red{Casiana: update}.
%
% \red{Bruce: add your favorite project code.}
%
AOG thanks Fundaci\'on La Caixa for financial support in the form of a PhD contract.
MEF is partly funded by the {\em Estallidos} project, as well as by the 
%M. E. F. gratefully acknowledges the financial support of 
{\em Fundação para a Ciência e Tecnologia} (FCT  Portugal), through the research grant SFRH/BPD/107801/2015.
JMA acknowledges  support from the European  Research Council Starting
Grant (SEDmorph; P.I. V. Wild).
%
%%%%%%%%%%%%%%%%%%%%%%%%%%%%
% 
%\newcommand\aj{AJ}
%\newcommand\apj{ApJ}
%\newcommand\apjl{ApJ}
%\newcommand\apjs{ApJS}
%\newcommand\mnras{MNRAS}
%\newcommand\aapr{A\&ARev}
%\newcommand\araa{ARA\&A}
%\newcommand\aap{A\&A}
%\newcommand\nat{Nature}
%\newcommand\pasp{PASP}
%\newcommand\nar{NewARev}
%\newcommand\na{NewA}
%\newcommand\pra{PhyRevA}
%\newcommand\ap{$\approx$}
%\newcommand\jcap{JCAP}
%Publications of the Astronomical Society of Australia
\newcommand\pasa{PASA}
%\newcommand\lt{{$<$}}
%\bibliography{ms}%letter,extra_ref}
\bibliographystyle{apj}%{aa}

%%%%%%%%%%%%%%%%%%%%%%%%%%%%%%%%%

\appendix
\section{Emission of an expanding dusty shell of gas}\label{appendix}

%\red{The eqs in this appendix are already corrected and ok.}

The emission of a uniform shell produces a top-hat line 
profile that is not consistent with the two-horned profiles observed in the  wings of H$\alpha$ 
(Fig.~\ref{fig_lobes}). However, if the shell has some internal extinction, 
then it preferentially absorbs photons emerging from the regions where 
the line-of-sight (LOS) tends to be parallel to the surface of the shell. Those 
are the regions of small Doppler shifts that contribute to the line center, 
thus creating a two-horned profile. 
\begin{figure}
\includegraphics[scale=0.7]{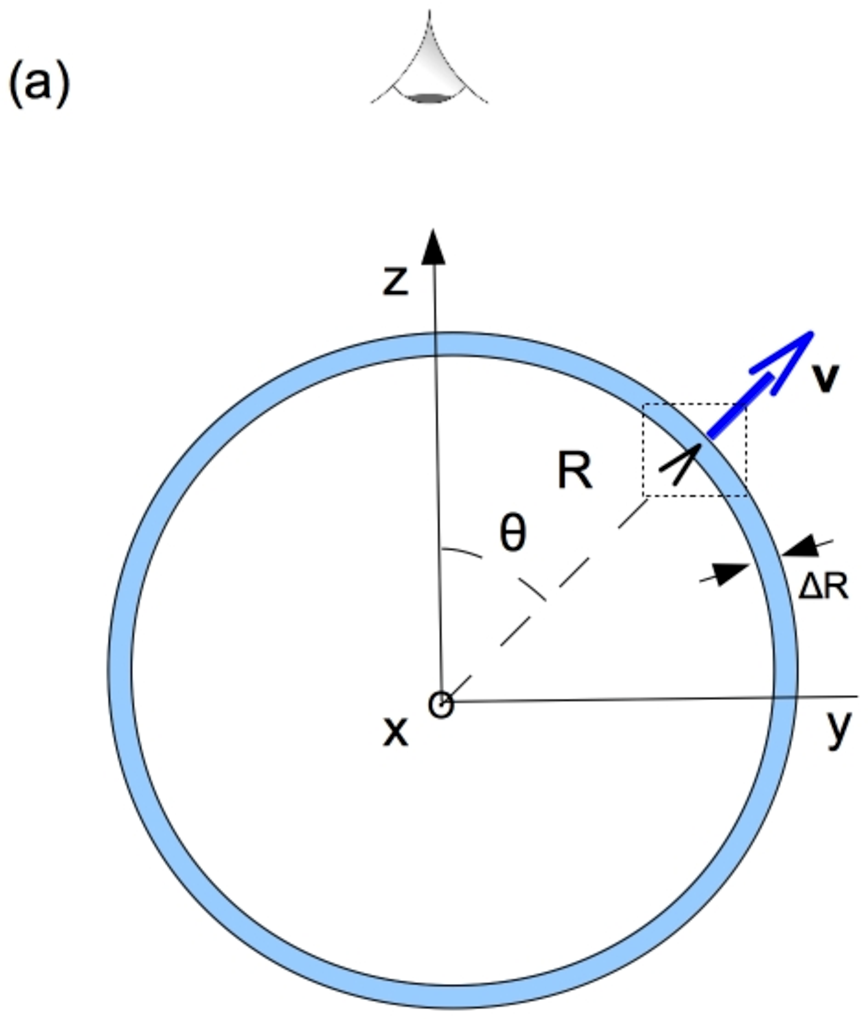}
\includegraphics[scale=0.6]{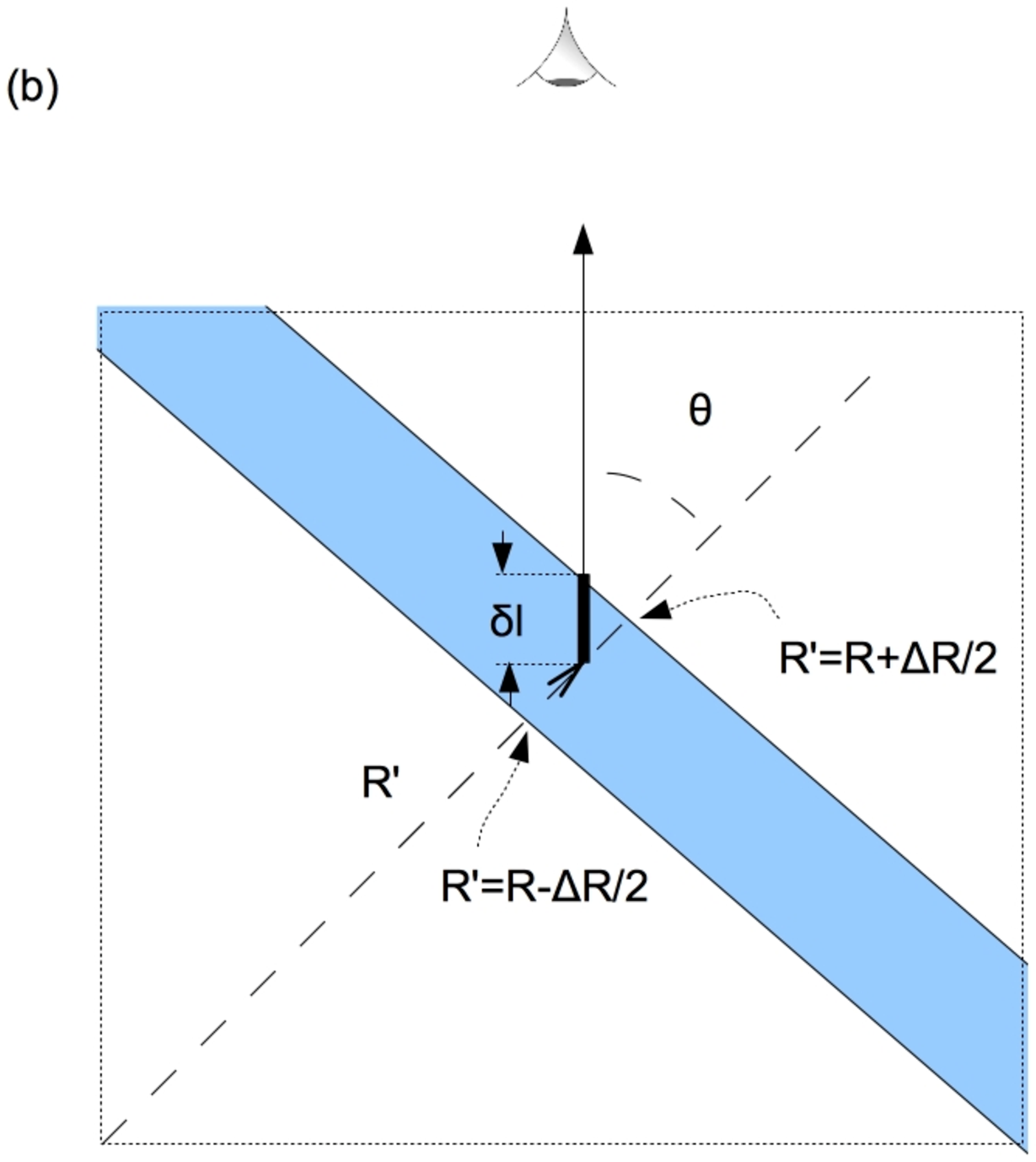}
\caption{Schematic used to compute the emission-line profile of an expanding dusty 
shell of radius $R$ and width $\Delta R$.
(a) Projection of the shell in the Y-Z plane, with Z in the direction along the
LOS. The velocity vector, ${\bf v}$, in the direction set by $\theta$ is shown as a 
blue solid-line arrow.
(b) Zoom-in of the region enclosed by the dotted-line rectangle in (a). 
The length along the LOS traversed by the photons emitted at coordinates ($\theta$,~$R'$)
%\red{is this correct?} 
is denoted as $\delta l$, with $\delta l=(R+\Delta R/2-R')/\cos\theta$.
}
\label{cartoon_shell}
\end{figure}

The effect can be readily shown assuming an expanding shell of radius $R$ and width $\Delta R$, 
as shown in Fig.~\ref{cartoon_shell}a. The velocity along the LOS, $v_z$, only depends on the
inclination angle, $\theta$,
\begin{equation}
v_z=v\,\cos\theta.
\label{app:eq1}
\end{equation}
Therefore,
the line profile, $I(v_z)$, is just the total energy emitted with velocity between $v_z$ and 
$v_z+\Delta v_z$, divided by $\Delta v_z$. Because of Eq.~(\ref{app:eq1}), it is proportional to the
energy emitted by the ring of the shell with inclination angles between $\theta$ and $\theta+\Delta\theta$,
i.e., 
\begin{equation}
\Delta s =\int_0^{2\pi}\int_{R-\Delta R/2}^{R+\Delta R/2}\int_\theta^{\theta+\Delta\theta}\,
\epsilon\, \sin\theta'\,R'^2\,d\phi\,d R'\, d\theta',
\label{app:eq2}
\end{equation}
where $R$ is the radius of the shell of width $\Delta R$, $\phi$ is the azimuth in spherical coordinates, 
and $\epsilon$ stands for the emission  per unit volume. Assuming the shell to be small enough ($\Delta R \ll R$), then
\begin{equation}
\Delta s \simeq 2\pi\, R^2\, \sin\theta\,\Delta\theta\, \int_{R-\Delta R/2}^{R+\Delta R/2}\epsilon\,dR'.
\label{app:eq3}
\end{equation}
According to the definition of $I(v_z)$, 
\begin{equation}
I(v_z)=\Big|{{\Delta s}\over{\Delta v_z}}\Big|=\Big|{{\Delta s}\over{\Delta\theta}}\,{{\Delta\theta}\over{\Delta v_z}}\Big|
\simeq\Big| {{\Delta s}\over{\Delta\theta}}\,
      \Big({{dv_z}\over{d\theta}}\Big)^{-1}\Big|,
\label{app:eq4}
\end{equation}
so Eqs.~(\ref{app:eq1}), (\ref{app:eq3}) and (\ref{app:eq4}) yield,
\begin{equation}
I(v_z)\simeq 2\pi\, R^2\,v^{-1}\,\int_{R-\Delta R/2}^{R+\Delta R/2}\epsilon\,dR',
\label{app:eq5}
\end{equation}
which holds only for $|v_z|\leq v$. Outside this range of velocities, $\Delta s$ is 
equal to zero, and so is $I(v_z)$.  

In the case that $\epsilon$ is constant, then $I(v_z)$ is also constant (Eq.~[\ref{app:eq5}]),
leading to the top-hat line profiles expected from an optically thin expanding shell.
We will consider the case where the emission is constant, $\epsilon_0$, but the 
shell absorbs part of the emitted photons due to internal extinction, $\beta$, i.e.,
\begin{equation}
\epsilon=\epsilon_0\,\exp(-\beta\delta l),
\label{app:extintion}
\end{equation}
with 
\begin{equation}
\delta l(R')\simeq {{R+\Delta R/2- R'}\over{\cos\theta}},
\label{app:eq6}
\end{equation}
as shown in Fig.~\ref{cartoon_shell}b. 
%The symbol $\beta$ stands for the extinction optical depth per unit length.
For the moment, we have
assumed that the emission occurs in the upper cap only. Using Eqs.~(\ref{app:eq5}), (\ref{app:extintion}), and (\ref{app:eq6}),
\begin{equation}
I(v_z)\simeq I_0\,{{\cos\theta}\over{\beta\Delta R}}\Big[1-\exp(-\beta\Delta R/\cos\theta)\Big],
\label{app:eq7}
\end{equation}
\begin{displaymath}
\cos\theta \geq 0, {\rm ~~~~~or~~~~~}v_z \geq 0,
\end{displaymath}
with $I_0$ the emitted intensity if there were no extintion ($\beta \longrightarrow 0$),
\begin{equation}
I_0=2\pi\,R^2\,\Delta R\,\,v^{-1}\,\epsilon_0.
\end{equation}

The case of the lower cap is similar, except that one has to consider the additional
absorption produced by the upper cap. In this other case, 
\begin{equation}
I(v_z)\simeq I_0\,
{{|\cos\theta|}\over{\beta\Delta R}}\Big[1-\exp(-\beta\Delta R/|\cos\theta|)\Big]\,\exp(-\beta\Delta R/|\cos\theta|),
\label{app:eq8}
\end{equation}
with
\begin{displaymath}
\cos\theta < 0 {\rm ~~~~~or~~~~~}v_z < 0.
\end{displaymath}

\begin{figure}
\centering
\includegraphics[scale=0.5]{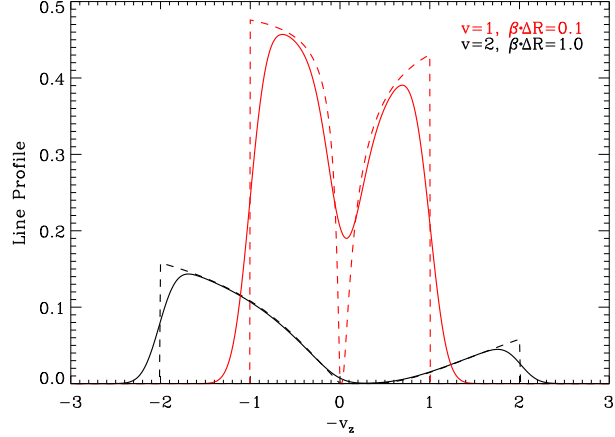}
\caption{Line profiles expected from an expanding dusty shell. 
They portray both weak internal extinction ($\beta\Delta R=0.1$; profiles
in red), and significant internal extinction ($\beta\Delta R=1.0$, profiles in black).
Red and black lines also differ in the expansion speed $v$, as shown in the inset. 
We include the profiles predicted by Eqs.~(\ref{app:eq7}) and (\ref{app:eq8}) (dashed lines), together
with these profiles convolved with a Gaussian to represent the finite spectral
resolution of the spectrograph (solid lines). 
We assume the FWHM of the Gaussian to be 0.35 in the dimensionless velocity scale. 
If this value is ascribed to the resolution of our observation ($\sim 40~$km~s$^{-1}$), 
then the expansion speeds for the shells giving rise to the red and black profiles 
are 110~km~s$^{-1}$ and 220~km~s$^{-1}$, respectively. 
Abscissae show $-v_z$ rather than $v_z$ so that the positive values correspond 
to redshifts.}
\label{app:example}
\end{figure}

The typical values of the extinction coefficient at H$\beta$ in XMPs are between 
$0.1$ and $0.9$ \citep{2016ApJ...819..110S}.  These values were inferred from the Balmer decrement in 
Sloan Digital Sky Survey (SDSS) spectra of XMPs. If we use them to represent the total extintion in the 
shell, then $0.15 < \beta\,\Delta R < 1.5$, where we have considered the differential extintion
between H$\alpha$ and H$\beta$ using the extinction law in the Milky Way by \citet{1989ApJ...345..245C}.
The dashed lines in Fig.~\ref{app:example} show examples of the line profiles 
resulting from Eqs.~(\ref{app:eq7}) and (\ref{app:eq8}).
They represent extinctions that cover the range of values mentioned above. 
We show the original profiles (dashed lines) plus their convolution with a Gaussian
that represents the finite resolution of the observation (solid lines). The width 
of the Gaussian has been chosen so that, if it represents our WHT observations
(FWHM~$\sim 40~$km~s$^{-1}$), then the expansion velocities of the 
shells are similar to the speeds of the multiple components analyzed
in Sect.~\ref{multiple_comp}. 
One can see the two-horned shape, with a redshifted lobe that is smaller 
than the blushifted lobe; the redshifted photons come from the lower cap 
of the shell, so that they are also extincted by the upper cap (we have flipped the sign of the abscissa axis 
in Fig.~\ref{app:example}, so that positive corresponds to redshift, as usual).

The model in Eqs.~(\ref{app:eq7}) and (\ref{app:eq8}) 
also predicts the fraction of emitted light lost by internal  
absorption. For the two profiles shown in Fig.~\ref{app:example}, the 
fraction of light that emerges is 74\,\% and 24\,\% of the emitted
light for the red and black profiles, respectively.

\end{document}